\documentclass[12pt]{article}
\usepackage[utf8]{inputenc}
\usepackage{graphicx}
\usepackage{amsmath}
\usepackage{amssymb}
\usepackage{amsfonts}
\usepackage{float}
\usepackage{listings}
\usepackage{caption}
\usepackage{color}
\usepackage{csquotes}
\usepackage[english]{babel}
\usepackage[affil-it]{authblk}
\usepackage[left=2.5cm,right=2.5cm, top=3cm, bottom=3cm]{geometry}
\usepackage{mathcomp}
\usepackage{breqn}
\usepackage{tabularx}
\usepackage{makecell}
\usepackage{hyperref}

\usepackage[backend=biber,bibencoding=utf8, style=numeric-comp,sorting=none]{biblatex}
\bibliography{Bibliography}
\hypersetup{
    pdftoolbar=true,        
    pdfmenubar=true,       
    pdffitwindow=false,     
    pdfstartview={FitH},
    pdfauthor={Parangam Goswami},     
    colorlinks=true,       
    linkcolor=red,         
    citecolor=blue,       
}
\makeatletter
\newbox\abstract@box
\renewenvironment{abstract}
  {\global\setbox\abstract@box=\vbox\bgroup
     \hsize=\textwidth\linewidth=\textwidth
    \small
    \begin{center}%
    {\bfseries \abstractname\vspace{-.5em}\vspace{\z@}}%
    \end{center}%
    \quotation}
  {\endquotation\egroup}
\expandafter\def\expandafter\@maketitle\expandafter{\@maketitle
  \ifvoid\abstract@box\else\unvbox\abstract@box\if@twocolumn\vskip1.5em\fi\fi}
\makeatother
\providecommand{\keywords}[1]{\textbf{\textit{Keywords---}} #1}
\begin{document}
\definecolor{dkgreen}{rgb}{0,0.6,0}
\definecolor{gray}{rgb}{0.5,0.5,0.5}
\definecolor{mauve}{rgb}{0.58,0,0.82}

\lstset{frame=tb,
  	language=Matlab,
  	aboveskip=3mm,
  	belowskip=3mm,
  	showstringspaces=false,
  	columns=flexible,
  	basicstyle={\small\ttfamily},
  	numbers=none,
  	numberstyle=\tiny\color{gray},
	commentstyle=\color{dkgreen},
  	stringstyle=\color{mauve},
  	breaklines=true,
  	breakatwhitespace=true
  	tabsize=3
}
    \title{Non-commutative Wormholes in {$f(R)$} Gravity satisfying the Energy Conditions}
    \author{Anshuman Baruah\thanks{E-mail: \textsf{anshuman.baruah@aus.ac.in}}}
    \author{Parangam Goswami\thanks{E-mail: \textsf{parangam.goswami@aus.ac.in}}}
    \author{Atri Deshamukhya\thanks{E-mail: \textsf{atri.deshamukhya@gmail.com}}}
    \affil{Department of Physics, Assam University, Cachar - 788011, Assam, India}
    \date{}
    \begin{abstract}
   Traversable wormholes in General Relativity (GR) require exotic matter sources that violate the null energy condition (NEC), and such behavior may be avoided in modified gravity. Moreover, the concept of non-commutative geometry as a gravitational source can be leveraged both in GR and modified gravity to realize non-trivial space-time configurations. In this study, we use $f(R)$ gravity in conjunction with non-commutative geometry to analyze spherically symmetric traversable Morris-Thorne wormhole solutions from the aspect of energy condition violation, considering both constant and variable red-shift functions. First, we use well constrained metric and model parameters in a viable $f(R)$ gravity model to demonstrate that wormholes respecting the NEC can be obtained with suitable choices of parameters. Additionally, we check the strong and dominant energy conditions to further validate our results. We then leverage non-commutative geometry in the framework of $f(R)$ gravity to show that wormholes respecting the different energy conditions with a phantom-like source can be realized with suitable choices of model parameters. Our comprehensive analyses using well-constrained model parameters show that wormholes satisfying the NEC can be realized in the framework of non-commutative geometry with modified gravity.
   \\
   \\
    \keywords{Wormhole; Energy Condition; f(R) gravity; Non-commutative geometry}
    \end{abstract}

    \maketitle

    \section{Introduction}
    \label{sec:int}
    Wormholes are space-time configurations with topologically simple boundaries and non-trivial interiors \cite{Visser:1995cc}, generally interpreted as tubes or bridges connecting two asymptotically flat regions of space-time. Wormholes can be obtained as exact solutions of the Einstein's field equations (EFEs) in general relativity (GR) \cite{Flamm:1916, Einstein:1935tc}, and lead to interesting physical implications. The most widely discussed of these is the prospect of traversable Lorentzian wormholes. By traversability, we mean that the throat of the wormhole is stable, and remains open for signals that fall through the first asymptotically flat region to emerge in the other region. Initial solutions such as that due to Ellis \cite{ellis1973ether}, were either unstable or geodesically incomplete, and the first stable and geodesically complete traversable wormhole solution was reported by Morris \& Thorne in 1988. The main highlight of their result is that such wormholes can only be constructed at the expense of violating the null energy condition (NEC). Morris-Thorne traversable wormholes are described by the following spherically symmetric line element \cite{Morris:1988cz}:
    \begin{equation}
     ds^2 = - e^{2 \Phi(r)} dt^2 + \frac{dr^2}{ 1 - \frac{b(r)}{r}} + r^2 {d{\theta}}^2 + r^2 sin^2 {\theta} d {\phi}^2
     \label{mtle}
\end{equation}
The first factor $e^{2 \Phi(r)}$ in the above line element is used to estimate the gravitational redshift following the standard approach in spherically symmetric space-times, and $\Phi(r)$ is referred to in literature as the redshift function. Moreover, $b(r)$ determines the topological configuration of the space-time, and is called the wormhole shape function. The wormhole throat is located at some value $r_0$ of the radial coordinate $r$, which is referred to as the throat radius. Traversability demands that the throat is not surrounded by an event horizion. Horizons in spherically symmetric space-times are identified by physically non-singular surfaces at $g_{00}=-e^{2\Phi}\rightarrow 0$, which leads to the constraint that $\Phi(r)$ must be well-defined throughout the space-time. Moreover, geometric restrictions on the throat demand that (i) $b(r_o) = r_o$, (ii) $\frac{b(r) - b^{\prime} (r) r}{b^2} > 0$, (iii) $b^{\prime} (r_o) - 1 \leq 0$, (iv) $\frac{b(r)}{r} < 1, \forall r > r_o$, (v) $\frac{b(r)}{r} \rightarrow 0$ as $r \rightarrow \infty$. These constraints on the metric functions constrain the allowed energy density ($\rho$), radial ($p_r$), and tangential ($p_t$) pressures of matter sources that can support traversable wormholes through the EFEs. The NEC and weak energy condition (WEC) require $\rho+p_r \geq 0$ and $\rho \geq 0$, which ensures that the energy density of the source is measured to be positive. By contrast, traversable wormholes in GR require $\rho+p_r < 0$, thus violating the NEC. This behavior is unphysical from the perspective of the Standard Model. However, traversable wormholes are an inherent feature of modified gravity theories as well, where such violations may be avoided. The gravitational field equations in modified/extended gravity can be recast as \cite{PhysRevD.80.104012,capo1, capo2}:
\begin{equation}
    g_1 (\Psi^i) (G_{\mu \nu} + H_{\mu \nu}) = 8 \pi g_2 ( \Psi^j )T_{\mu \nu}
\end{equation}
where $H_{\mu \nu}$ comprises geometrical corrections to GR in modified gravity, $g_i (\Psi^i)$ are multiplicative factors, and $\Psi^i$ are curvature invariants or other fields contributing to the dynamics of the theory. Accurately identifying $g(\Psi^j) $ and $H_{\mu \nu}$ makes it possible to formulate wormhole solutions in a manner such that the matter stress energy obeys the corresponding NEC ($T_{\mu \nu}k^\nu k^\nu \geq 0$), and violations may be attributed to additional curvature terms. For example, the generalized NEC for $f(R)$ gravity can be written as \cite{capo1, capo2} $T_{\mu \nu} k^\mu k^\nu + k^\mu k^\nu \frac{\nabla_\mu \nabla_\nu F}{F} \geq 0$, with $F \equiv d f(R)/dR$. In such a setting, an appropriate form of $f(R)$ may be considered such that the above inequality holds, even when $T_{\mu \nu}k^\nu k^\nu \geq 0$. Thus, owing to the inherently different structure of the field equations in modified gravity, wormholes may be realized with matter sources satisfying the energy conditions, with violations attributed to additional degrees of freedom in the theory. This interesting feature has been discussed extensively in literature, especially in the context of $f(R)$ modified gravity.
\\
$f(R)$ theories modify the Hilbert action in GR by substituting the Ricci scalar $R$ with some arbitrary function of it. This simple modification can account for the shortcomings of GR, such inflation and late-time expansion of the observable Universe. It is well-known that wormholes can be obtained with ordinary matter sources in modified gravity theories \cite{PhysRevD.80.104012,pavlovic2015wormholes,PhysRevD.96.044038,PhysRevD.97.024007}. Field equations for traversable wormhole solutions in metric $f(R)$ gravity have been reported \cite{PhysRevD.80.104012,pavlovic2015wormholes}, and solutions have been discussed in models such as $f(R)=1/R$, $f(R)=R^2$, $f(R) = R + \alpha R^m - \beta R^{-n}$ \cite{Furey_2004,2019D}. Evolution of cosmological wormhole solutions in $f(R)$ gravity suggested that exotic matter is required near the wormhole throat only in the limit of $f(R) = R$ \cite{PhysRevD.94.044041}. Wormhole solutions and the energy conditions in the framework of $f(R)$ and $f(R,T)$ modified gravity have been studied with different choices of the shape function such as exponential, logarithimic and trigonometric \cite{pavlovic2015wormholes,doi:10.1142/S0218271820500686,doi:10.1142/S0218271819500391,azizi2013wormhole,shweta2020traversable,2019A,2020C}. Moreover, it is worth noting here that several recent developments have been made in the field of modified gravity in the form of novel models such as $f(Q)$ and $f(T)$ gravity, and wormhole solutions have been discussed extensively in both $f(Q)$ \cite{fq,fq2,fq3,fq4} and $f(T)$ gravity \cite{ft1,ft2,ft3}. 
\\
The inherent structure of space-time has also been discussed significantly in the context of modifications of GR, and the concept of non-commutative geometry is of particular interest \cite{banerjee2009topics}. String theory suggests that there is a lower bound to distance measurement, and at small length scales, co-ordinates behave as non-commutative operators on a D-Brane discretizing space-time \cite{witten1995string, seiberg1999string}. This approach discretizes space-time with a commutator of the form $[\textbf{x}^\mu, \textbf{x}^\nu = i \theta^{\mu \nu}]$, where $\theta^{\mu \nu}$ is an antisymmetric tensor \cite{snyder1947quantized}. An interesting physical implication of this is that at small length scales, particles become smeared objects modelled with a Gaussian or Lorentzian distribution. The geometric contribution thereof can be modelled as a self-gravitating source of anisotropic fluid with an energy density profile determined uniquely by $\theta^{\mu \nu}$ and the effective mass \cite{Nicolini_2009,rinaldi2011new}. Non-commutative geometry as a gravitational source is an intrinsic property of space-time, and does not depend on curvature \cite{NICOLINI2006547}. To this end, wormhole solutions with viable physical properties in non-commutative backgrounds have been discussed previously \cite{rahaman2015wormhole, rahaman2012searching}. Moreover, self-consistent wormhole solutions in the context of non-commutative geometry in the semi-classical limit have also been proposed in $f(R)$ and $f(R,T)$ modified gravity \cite{GARATTINI2009146,zubair2017existence}. In this study, we leverage $f(R)$ gravity to investigate possible wormhole solutions with and without invoking a non-commutative background, especially focusing on energy condition violations. Our results demonstrate that wormholes satisfying the NEC can be obtained in the non-commutative background supported by a phantom-like matter sources. Furthermore, we analyze the stability of the wormhole space-time using the generalized Tolman-Oppenheimer-Volkov (TOV) equation, and also comment on the amount of the exotic matter using the volume integral quantifier (VIQ). 
\\
The remainder of this manuscript is organized as follows. In Section \ref{sec:whgeometry}, we describe the basic features of Morris-Thorne wormholes in $f(R)$ gravity, introduce the models studied, and setup the field equations. In Section \ref{sec:Res}, we present numerical analyses of the energy conditions, and discuss the implications of these results. Finally, we conclude with a few remarks in Section \ref{sec:Con}. Throughout the work, we adhere to the natural system of units ($G=c=1$). 

\section{Wormholes in $f(R)$ gravity and non-commutative geometry}\label{sec:whgeometry}

 \subsection{Wormholes in $f(R)$ gravity}
   $f(R)$ theories are described by a general action of the form:
    
    \begin{equation}
        S = \int d^4 x \sqrt{-g} \left[ f(R) + \mathcal{L}_m \right]
        \label{fract}
    \end{equation}
    
    where the symbols imply their usual meanings. The modified EFE for $f(R)$ gravity in the metric formalism can be obtained as:
    
    \begin{equation}
    FR_{\mu\nu}-\frac{1}{2}f(R)\,g_{\mu\nu}-\nabla_\mu \nabla_\nu
    F+g_{\mu\nu}\Box F=\,T^m_{\mu\nu} \,,
    \label{eq:3}
    \end{equation}
    
    where $F\equiv df/dR$, and $T^m_{\mu\nu} = \frac{-2}{\sqrt{-g}} \frac{\delta \mathcal{L}_m}{\delta g^{\mu \nu}}$ denotes the matter stress-energy tensor. We consider that $g_{\mu\nu}$ describes a spherically symmetric space-time described by the line-element in Eq. (\ref{mtle}), and contract Eq. (\ref{eq:3}) to yield the following:
\begin{equation}
FR-2f(R)+3\Box F=T
\label{eq:4}
\end{equation}
 
Here, $T$ is the trace of the matter stress energy tensor and $\Box F$ is given by:
\begin{align}
    \Box F= \frac{1}{\sqrt{-g}} \partial_\mu (\sqrt{-g} g^{\mu \nu} \partial_\nu F) = \left(1-\frac{b}{r}\right)\left[F''
-\frac{b'r-b}{2r^2(1-b/r)}\,F'+\frac{2F'}{r}\right]
\end{align}
with $F'=d F/d R$ and $b'=d\,b(r)/dr$. Now, substituting Eq. (\ref{eq:4}) in Eq. (\ref{eq:3}) yields the following modified EFE:
\begin{equation}
G_{\mu\nu}\equiv R_{\mu\nu}-\frac{1}{2}g_{\mu\nu} R= T^{{\rm
eff}}_{\mu\nu} \,
    \label{eq:6}
\end{equation}
Here, $T^{{\rm eff}}_{\mu\nu}$ is an effective stress-energy tensor, generally interpreted as a gravitational fluid responsible for NEC violations. $T^{{\rm eff}}_{\mu\nu}$ comprises the matter stress energy tensor $T^m_{\mu\nu}$ and curvature stress-energy tensor $T^{{\rm c}}_{\mu\nu}$ given by:
\begin{align}
T^{c}_{\mu\nu}=\frac{1}{F}\left[\nabla_\mu \nabla_\nu F
-\frac{1}{4}g_{\mu\nu}\left(RF+\Box F+T\right) \right]
    \label{eq:7}
\end{align}
We consider an anisotropic distribution of matter threading the wormhole geometry: 
\begin{equation}
T_{\mu\nu}=(\rho+p_t)U_\mu \, U_\nu+p_t\,
g_{\mu\nu}+(p_r-p_t)\chi_\mu \chi_\nu \,,
\end{equation}
where $U^\mu$ is the four-velocity, and $\chi^\mu$ is a unit space-like vector.
\\
In the background Eq. (\ref{mtle}), the following several expressions can be obtained \cite{PhysRevD.80.104012},
\begin{align}
\rho=\frac{Fb'}{r^2}
\label{generic1}
\end{align}
\begin{align}
p_r=-\frac{bF}{r^3}+\frac{F'}{2r^2}(b'r-b)-F''\left(1-\frac{b}{r}\right)
\label{generic2}
\end{align}
\begin{align}
p_t=-\frac{F'}{r}\left(1-\frac{b}{r}\right)+\frac{F}{2r^3}(b-b'r)
\label{generic3}
\end{align}


Here, $R= \frac{2b'}{r^2}$, and $\Phi'(r) = 0$ is generally considered for simplicity in calculations. However, $\Phi(r)$ has a crucial role in determining the energy conditions, as can be inferred from the expressions given below, and assuming $\Phi'(r) = 0$ might be an oversimplification. Thus, we also consider wormhole solutions with a non-constant redshift function ($\Phi'(r) \neq 0$) in our analyses. As will be highlighted later from the profile of the strong energy condition (SEC), this approach leads to a physically well-motivated solution. With $\Phi'(r) \neq 0$, we have 

\begin{equation}
    R=\left(1-\frac{b(r)}{r}\right) \left(2 \Phi ''(r)+4 \Phi '(r)^2+\frac{4 \Phi '(r)}{r}\right)-\frac{r b'(r)-b(r)}{r^2}\Phi '(r)-\frac{2 b'(r)}{r^2}
\end{equation}

Then, we have the following general expressions:
\begin{equation}\label{vrsfone}
    \rho=\frac{Fb'(r)}{r^2}-\Bigg(1-\frac{b(r)}{r}\Bigg)F'\Phi'(r)-H
\end{equation}
\begin{align}
    p_r&=-\frac{b(r)F}{r^3}+2\Bigg(1-\frac{b(r)}{r}\Bigg)\frac{\Phi^{'}(r)F}{r}
    &-\Bigg(1-\frac{b(r)}{r}\Bigg)\Bigg[F''+\frac{F'(rb'(r)-b(r))}{2r^2\Big(1-\frac{b(r)}{r}\Big)}\Bigg]+H
\end{align}
\begin{align}\label{vrsfthree}
p_t=\frac{F(b(r)-rb'(r))}{2r^3}-\frac{F'}{r}\Bigg(1-\frac{b(r)}{r}\Bigg)
&+F\Bigg(1-\frac{b(r)}{r}\Bigg)\\ \nonumber\Bigg(\Phi^{''}(r)-\frac{(rb'(r)-b(r))\Phi^{'}(r)}{2r(r-b)}  &+\Phi^{'2}(r)+\frac{\Phi^{'}(r)}{r}\Bigg)+H
\end{align}
where, $H(r)=\frac{1}{4}\left(FR+\Box F +T\right)$, for notational simplicity.\\

    \textit{The energy conditions}: In field theories such as GR and its modifications, the energy conditions \cite{curiel2017primer} are a set of relations that the matter stress energy tensor must satisfy, so that the energy density of the matter fields is measured to be positive by any observer traversing a time-like curve. The NEC requires that $\rho+p_r \geq 0$ and $\rho+p_t \geq 0 \, \forall r > 0$. The WEC requires that $\rho \geq 0 \, \forall r > 0$, in addition to the NEC; the SEC requires that $\rho+p_r+2p_t \geq 0 \, \forall r \geq 0$, and the dominant energy condition (DEC) requires that $\rho-|p_r| \geq 0 $ and $\rho-|p_t| \geq 0 \, \forall r \geq 0$.
    \\
    Within this framework, the energy condition inequalities can be probed numerically by using suitable choices of model and metric parameters. It is evident from Eq. (\ref{generic1}) that $f(R)$ may be obtained analytically by fixing $b(r)$ and integrating. However, the obtained $f(R)$ model may not be of physical significance (cosmological and weak field viability) \cite{guo2014solar}. Thus, we leverage a well-studied cosmologically viable $f(R)$ gravity model \cite{amendola2010dark}: 
    
    \begin{equation}
    \label{fR}
        f(R) = R - \mu R_c {\left( \frac{R}{R_c} \right)}^p
    \end{equation}
    Here, $\mu, R_c > 0$, and $0<p<1$. For any $f(R)$ gravity model to be cosmologically viable, it has to satisfy the following conditions \cite{guo2014solar,amendola2010dark,PhysRevD.75.083504}: (i) $F>0$ for $R \geq R_{0}$, where, $F=\frac{df}{dR}$, and $R_{0}$ is the Ricci scalar at the present epoch. (ii) $\frac{dF}{dR} > 0$ for $R>R_{0}$, this condition signifies stability against cosmological perturbations, and the presence of a matter-dominated epoch. It is also required for consistency with local gravity tests. (iii) $f(R) \rightarrow R-2\Lambda$ for $R>>R_{0}$; this is required for consistency with local gravity tests, and for the presence of a matter-dominated epoch. (iv) $0<m<1$ at $r=-2$, where $m=\frac{R\frac{dF}{dR}}{F}$ and $r=-\frac{RF}{f}$, this condition is required for the stability of the late-time de-Sitter point. These conditions constrain the parameters as $p<3 \times 10^{-10}$ and $R_{c} = 10^{-29}$, so that the model remains cosmologically viable \cite{2019A,guo2014solar,amendola2010dark}. \\
    
    In addition to the energy conditions, we analyze two more parameters: the equation of state (EoS) parameter, $\omega = p_{r}/\rho$, and anisotropy parameter, $\Delta = p_{t}-p_{r}$. The EoS parameter encodes the nature of matter sources, and the anisotropy parameter quantifies if the geometry is attractive or repulsive \cite{Morris:1988cz}. Furthermore, we adopt well constrained choices of $b(r)$ and $\Phi(r)$ satisfying the constraints described in Section \ref{sec:int}.\\
    
    \textit{Stability of wormholes and the amount of exotic matter:} The generalized TOV equation provides information regarding the stability of the wormhole space-time \cite{oppenheimer1939massive,gorini2008tolman,Kuhfittig:2020fue,ponce1993limiting}, and is given by,
    \begin{equation}\label{eq:tov}
    -\frac{dp_{r}}{dr}-\frac{\epsilon^{'}(r)}{2}(\rho+p_{r})+\frac{2}{r}(p_{t}-p_{r})=0,
    \end{equation}
    where $\epsilon(r)=2\Phi(r)$. $F_h$ represents the hydrostatic force, $F_g$ the gravitational force, and $F_a$ the anisotropic force, respectively. The equilibrium anisotropic mass distribution is determined by these three terms of the TOV equation \cite{ponce1993limiting}. 
    \begin{equation}\label{stabcomp}
    F_{\mathrm{h}}=-\frac{dp_{r}}{dr},\;\;\;\;\;\;\;\;F_{\mathrm{a}}=\frac{2}{r}(p_{t}-p_{r}), \;\;\;\;\;\;\;\;F_{\mathrm{g}}=-\frac{\epsilon^{'}}{2}(\rho+p_{r})
    \end{equation}
    Thus, we check the corresponding terms of the TOV equation in order to probe the stability of the wormhole configurations.
    \\
    Detailed information about the violations of the energy conditions along the radial distance can be achieved through the averaged null energy condition $\int_{\lambda_1}^{\lambda_2} T_{ij}k^ik^j d \lambda \geq 0$. However, as it is only a line integral, a more generalized description of the amount of matter violating the energy conditions can be achieved by using a volume integral, called the volume integral quantifier (VIQ) \cite{visser2003traversable,kar2004quantifying,lobo2013new}, which is defined as,
    \begin{equation}
    \label{VIQ}
    I_v = \oint [\rho+p_r] dV = 8\pi\int_{r_0}^{a}(\rho+p_r)r^2 dr
    \end{equation}
    Assuming that an exterior metric is matched with the wormhole space-time with the stress-energy tensor cutting off at some $r=a$, the VIQ implies the amount of NEC violating matter required for the wormhole configuration.
    As $a \rightarrow r_0$, $I_v \rightarrow 0$ signifies that arbitrarily small quantities of NEC violating matter is required for the wormhole \cite{visser2003traversable,kar2004quantifying}.
    \\
    With these insights we consider the following cases for our analyses in this study:
    
    \subsubsection{Case I}
    Here, we use $b(r) = r_0 \log \left( \frac{r}{r_0} \right) + r_0$, reported previously in Ref. \cite{pavlovic2015wormholes}, to analyze wormholes in $f(R)$ gravity. The energy density $\rho$, radial pressure $p_{r}$, and transverse pressure $p_{t}$ are obtained from Eqs. (\ref{generic1} - \ref{generic3}). These expressions have been included in the Appendix \ref{AppendixA}.
    
    \subsubsection{Case II}
    Here, $b(r) = r_0 \log \left( \frac{r}{r_0} \right) + r_0$ and $\Phi(r)=\sqrt{\frac{r_0}{r}}$, the energy density $\rho$, radial pressure $p_{r}$, and transverse pressure $p_{t}$ are obtained from Eqs. (\ref{vrsfone} - \ref{vrsfthree}). These expressions are rather lengthy, and have been included in Appendix \ref{AppendixA} :
    
    \subsection{Non-commutative geometry}
    As described previously, non-commutative geometry replaces point like particles by smeared objects, and space-time is encoded in the commutator $[\textbf{x}^\mu, \textbf{x}^\nu = i \theta^{\mu \nu}]$. Here, $\theta^{\mu \nu}$ is an antisymmetric tensor discretizing space-time in a manner analogous to $\hbar$ in phase space. The energy density of a static and spherically symmetric smeared gravitational source can be modelled using a Lorentzian distribution \cite{hamid2012entropic,Rahaman:2013qza}:
    
    \begin{equation}
    \rho=\frac{M\sqrt{\beta}}{\pi^2(r^2+\beta)^2}
    \label{eq:22}
    \end{equation}
    where, $M$ is the mass of the centralized diffused object and $\beta$ is a non-commutative parameter. Essentially, the mass is considered to be diffused throughout a linear region $\sqrt{\beta}$. We consider this energy density as the source in the field equations, derive an $f(R)$, and check the energy condition inequalities for the logarithmic shape function. 
    
    \subsubsection{Case III}
    
    Here, the form of the shape function is $b(r) = r_0 \log \left( \frac{r}{r_0} \right) + r_0$ and for simplicity in the calculations we have considered a constant red-shift function ($\Phi'(r) =0$). Now, substituting Eq. (\ref{eq:22}) in Eq. (\ref{generic1}), we get:
    
    \begin{equation}
    \label{eq:Fr}
    F(r)=\frac{\sqrt{\beta } M r^3}{\pi ^2 r_0 \left(\beta +r^2\right)^2}
    \end{equation}
    
    The radial pressure $p_r$, and the transverse pressure $p_t$ are obtained from Eqs. (\ref{generic2}-\ref{generic3}). These expressions have been included in Appendix \ref{AppendixA}.\\
    
    \section{Results and Discussions}
    \label{sec:Res}
    With $R_c = 10^{-29}$, $\mu = 0.10$, $p=3 \times 10^{-11}$, and $r_{0} = 0.9$, we analyze the energy conditions, TOV equation, and VIQ for Case I (constant red-shift function), and Case II (variable red shift function) respectively. Then, we analyze the same for Case III, i.e., for the wormhole solution in a non-commutative background.\\
    
\begin{figure}[H]
        \includegraphics[width=\columnwidth]{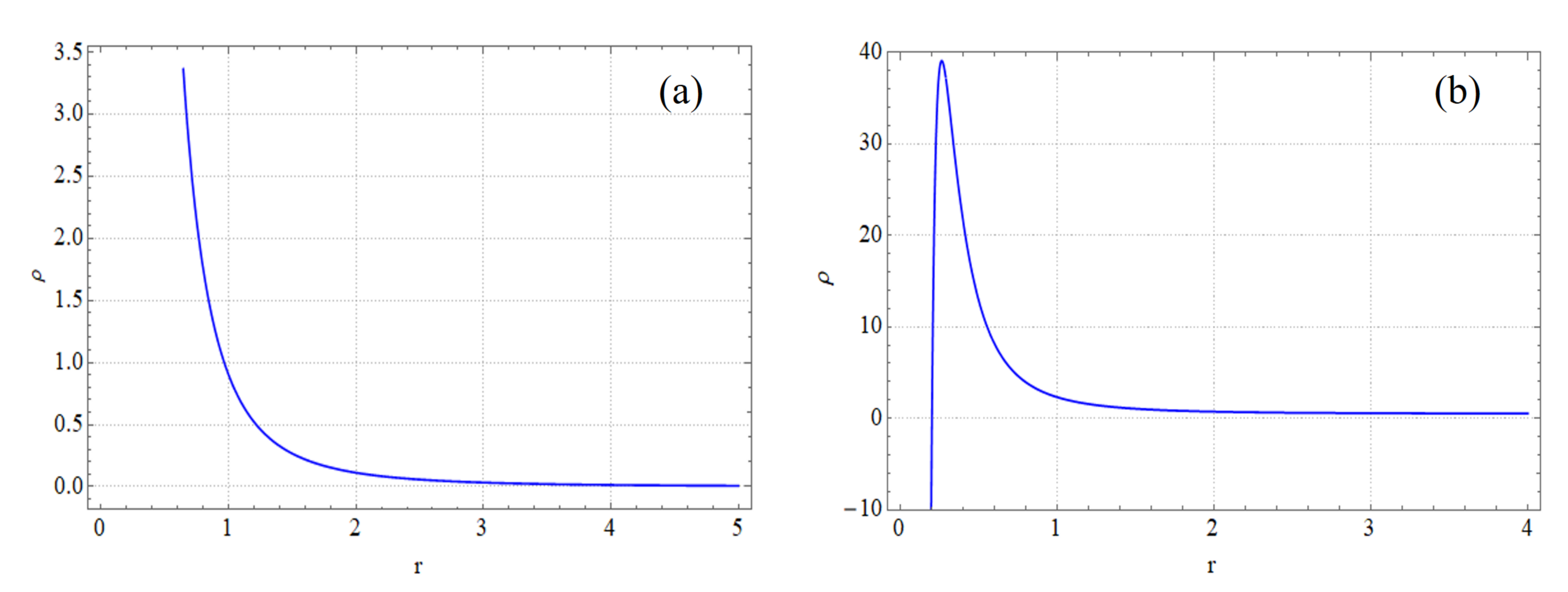}
        \caption{\centering
        Profile of the energy density $\rho$ vs. $r$ with $R_c = 10^{-29}$, \, $\mu = 0.10$, \, $p=3 \times 10^{-11}$, \, $r_{0} = 0.9$ for (a) Case I and (b) Case II.}
        \label{fig:one}
    \end{figure}

\begin{figure}[H]
        \includegraphics[width=\columnwidth]{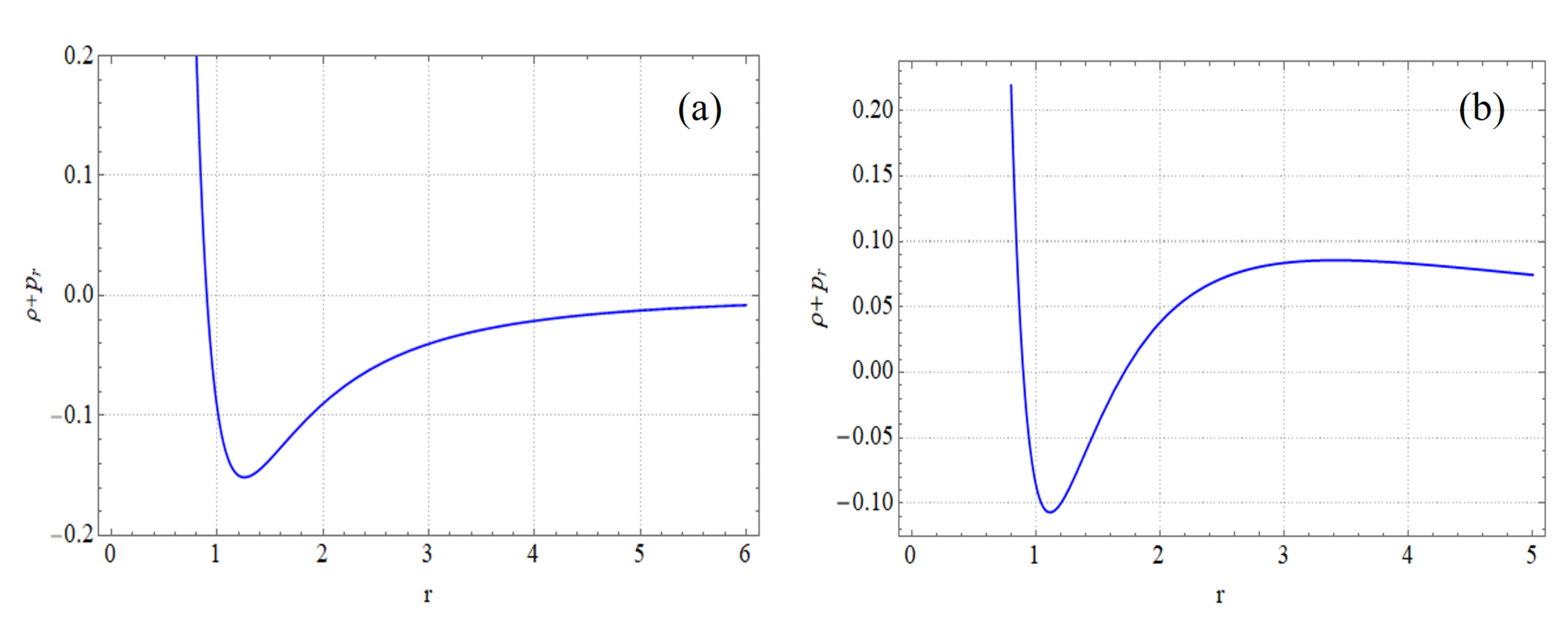}
        \caption{\centering Profile of the NEC term $\rho+p_{r}$ vs. $r$ with $R_c = 10^{-29}$, \, $\mu = 0.10$, \, $p=3 \times 10^{-11}$, \, $r_{0} = 0.9$ for (a) Case I and (b) Case II.}
        \label{fig:two}
    \end{figure}

\begin{figure}[H]
        \includegraphics[width=\columnwidth]{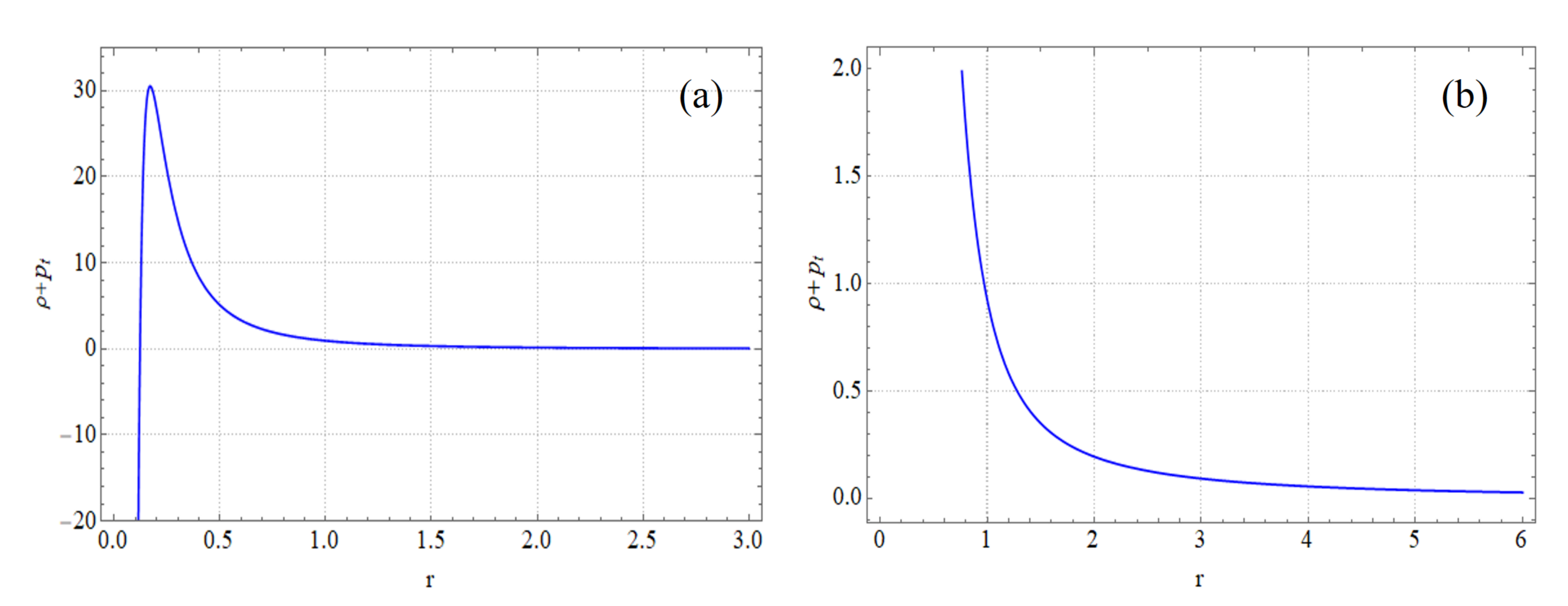}
        \caption{\centering Profile of the NEC term $\rho+p_{t}$ vs. $r$ with $R_c = 10^{-29}$, \, $\mu = 0.10$, \, $p=3 \times 10^{-11}$, \, $r_{0} = 0.9$ for (a) Case I and (b) Case II.}
        \label{fig:three}
    \end{figure}

    Figure \ref{fig:one} shows the plot of the WEC vs. radial distance $r$ in arbitrary units for both Case I and Case II, and it can be seen that the WEC is satisfied for both cases. Next, we analyze the NEC terms $\rho+p_{r}$, and $\rho+p_{t}$ for the two cases as shown in Figure \ref{fig:two} and Figure \ref{fig:three}, respectively. The plots of the first NEC term $\rho+p_{r}$ vs. radial distance $r$ are shown in Figure \ref{fig:two}(a) for Case I, and Figure \ref{fig:two}(b) for Case II, respectively.\\

\begin{figure}[H]
        \includegraphics[width=\columnwidth]{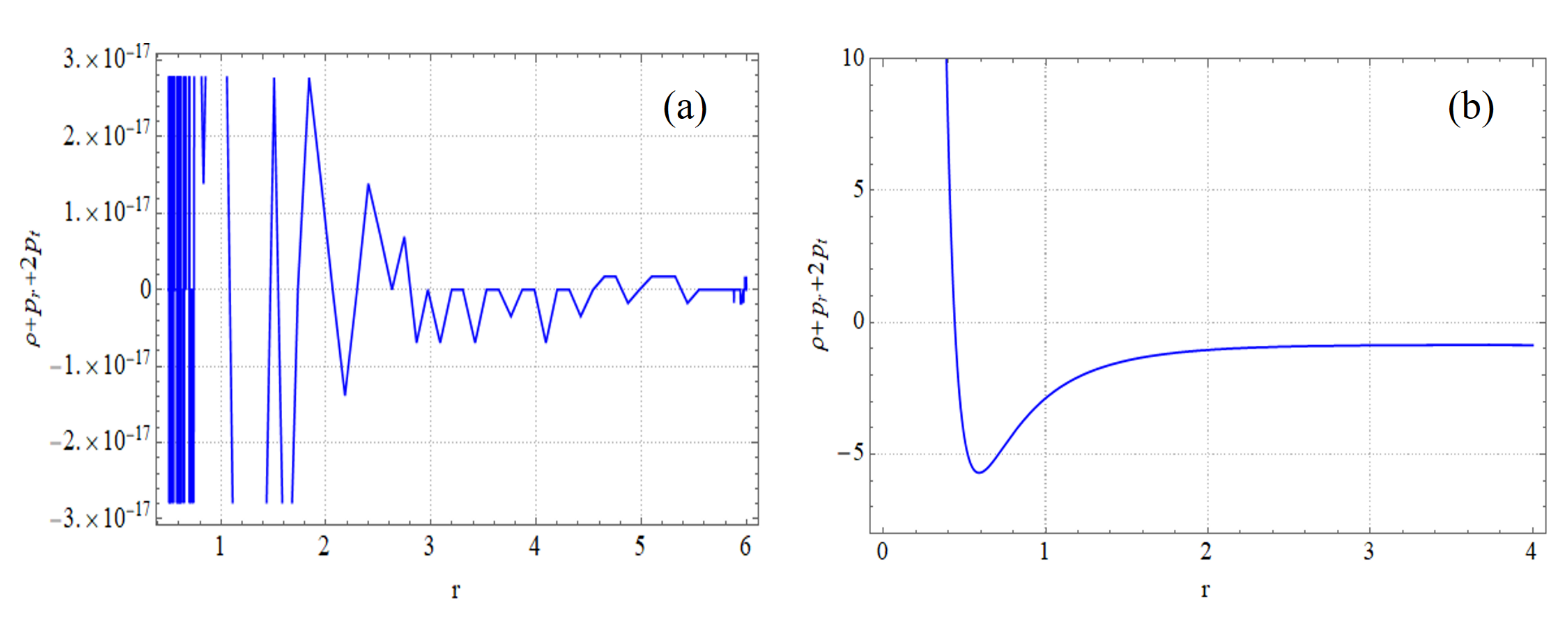}
        \caption{\centering Profile of the SEC $\rho+p_{r}+2p_{t}$ vs. $r$ with $R_c = 10^{-29}$, \, $\mu = 0.10$, \, $p=3 \times 10^{-11}$, \, $r_{0} = 0.9$ for (a) Case I and (b) Case II.}
        \label{fig:four}
    \end{figure}

    Figure \ref{fig:two} shows that the first NEC term is satisfied at the throat for both cases. Similarly, Figure \ref{fig:three} shows the plots of the second NEC term $\rho+p_{t}$ vs. the radial distance $r$ for (a) Case I and (b) Case II, and it can be inferred that the second NEC term is also satisfied near the throat for both cases. Next, we analyze the SEC term $\rho+p_{r}+2p_{t}$, and the plots are shown in Figure. \ref{fig:four}. From Figure \ref{fig:four}(a), it can be seen that for Case I the SEC oscillates between positive and negative values and becomes positive at large $r$. Thus, the SEC is inferred to be indeterminate at the throat and satisfied asymptotically. However, with the introduction of a variable red-shift function with the form $\Phi(r)=\sqrt{\frac{r_0}{r}}$ , Figure \ref{fig:four}(b) shows that the SEC is violated at the throat for Case II.\\

\begin{figure}[H]
        \includegraphics[width=\columnwidth]{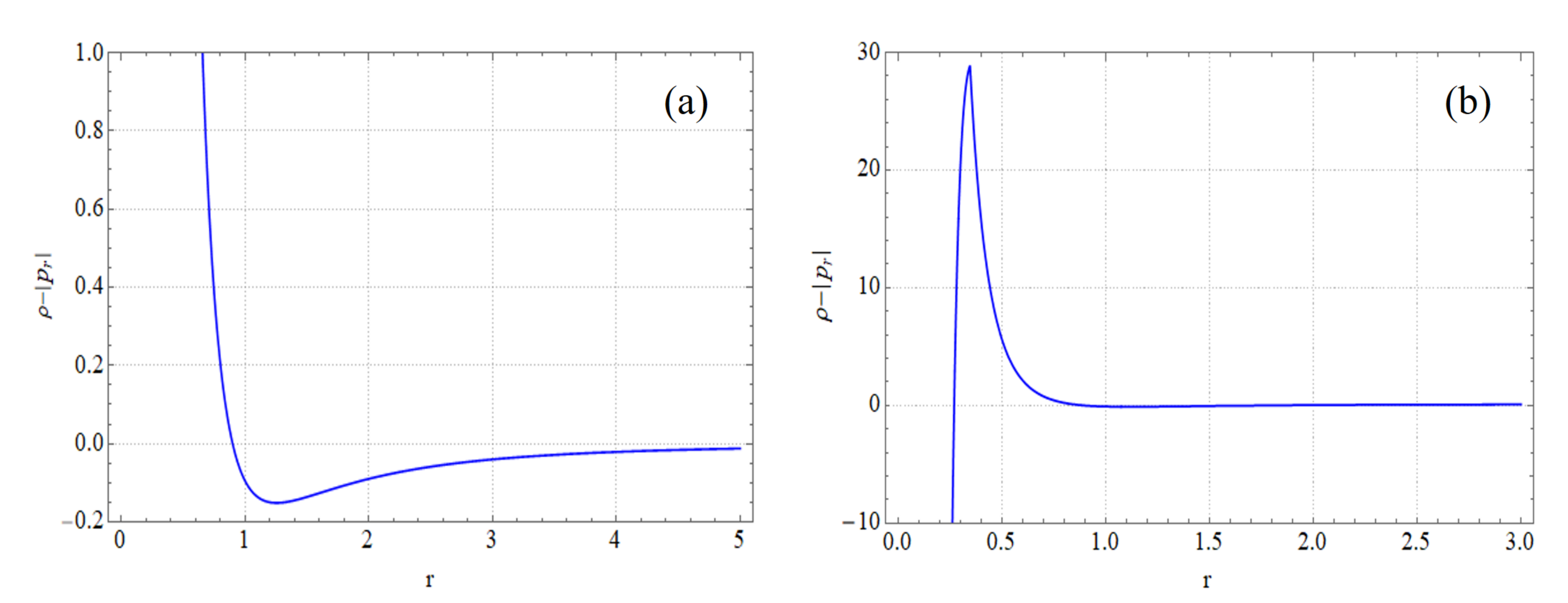}
        \caption{\centering Profile of the DEC term $\rho-|p_{r}|$ vs. $r$ with $R_c = 10^{-29}$, \, $\mu = 0.10$, \, $p=3 \times 10^{-11}$, \, $r_{0} = 0.9$ for (a) Case I and (b) Case II.}
        \label{fig:five}
    \end{figure}

\begin{figure}[H]
        \includegraphics[width=\columnwidth]{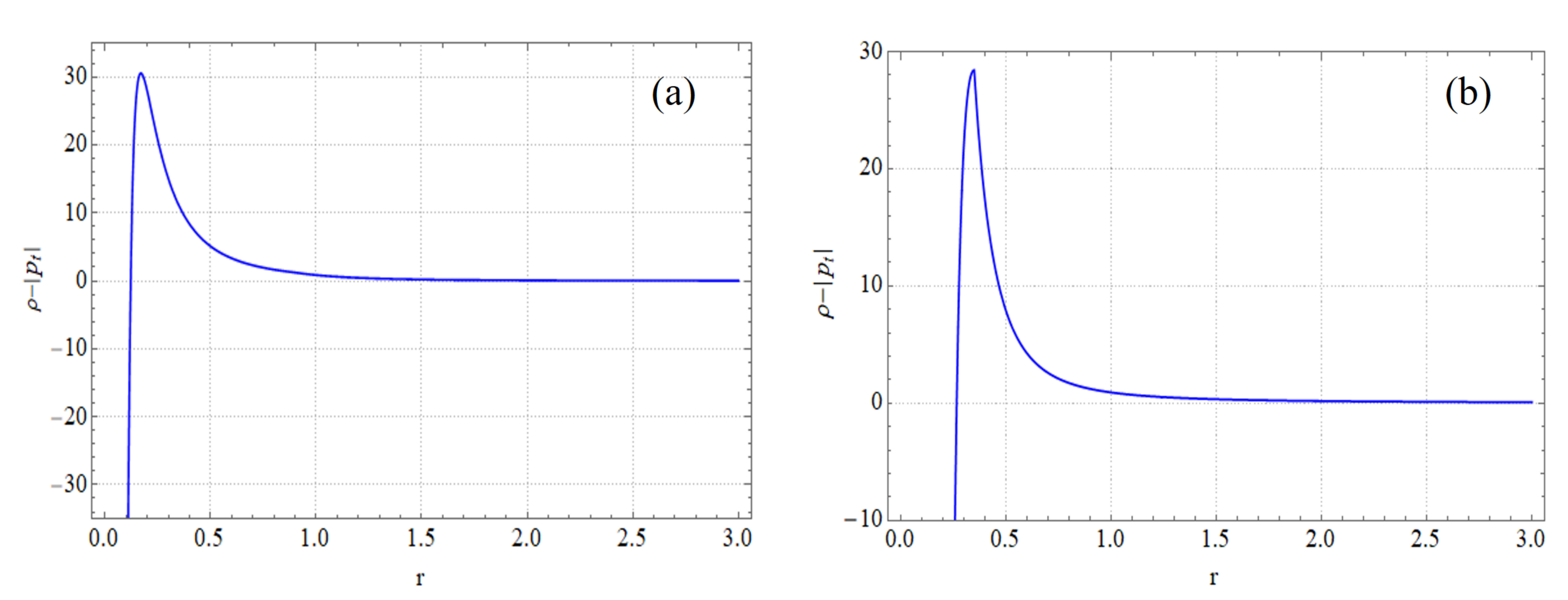}
        \caption{\centering Profile of the DEC term $\rho-|p_{t}|$ vs. $r$ with $R_c = 10^{-29}$, \, $\mu = 0.10$, \, $p=3 \times 10^{-11}$, \, $r_{0} = 0.9$ for (a) Case I and (b) Case II.}
        \label{fig:six}
    \end{figure}

    It is well known that violation of the SEC leads to late-time cosmic acceleration in modified gravity \cite{moraes2017simplest, albareti2013non}. Thus, the result in Case II highlights the importance of the redshift function in obtaining physically well-motivated solutions, as seen from the asymptotic behavior of the SEC in Figure \ref{fig:four}(b). The DEC terms $\rho-|p_r|$ and $\rho-|p_t|$ are shown in Figure \ref{fig:five} and Figure \ref{fig:six} respectively. Figure \ref{fig:five} shows that the first DEC term is satisfied at the throat for both Cases I and II. From Figure \ref{fig:six}, it can be seen that the second DEC term is also satisfied at the throat for both the cases. Next we analyze the EoS and the anisotropy parameters. \\

\begin{figure}[H]
        \includegraphics[width=\columnwidth]{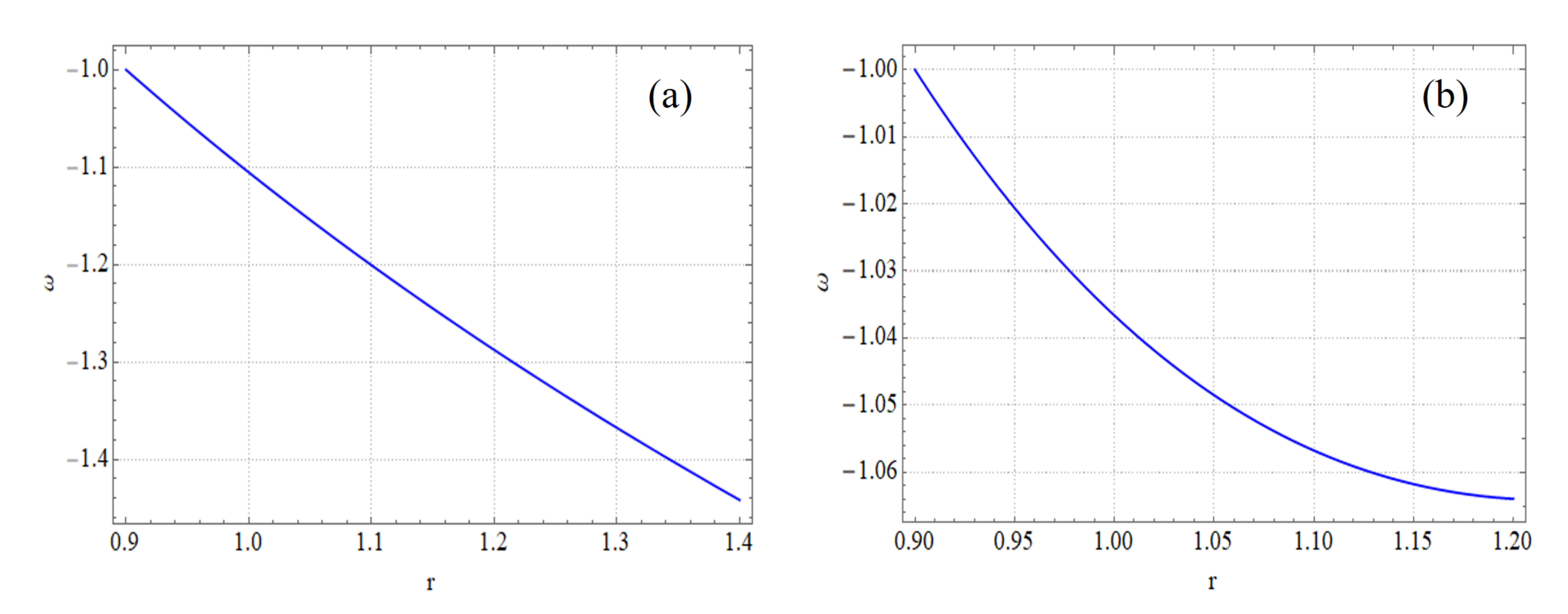}
        \caption{\centering Profile of the EoS parameter $\omega$ vs. $r$ with $R_c = 10^{-29}$, \, $\mu = 0.10$, \, $p=3 \times 10^{-11}$, \, $r_{0} = 0.9$ for (a) Case I and (b) Case II.}
        \label{fig:seven}
    \end{figure}

\begin{figure}[H]
        \includegraphics[width=\columnwidth]{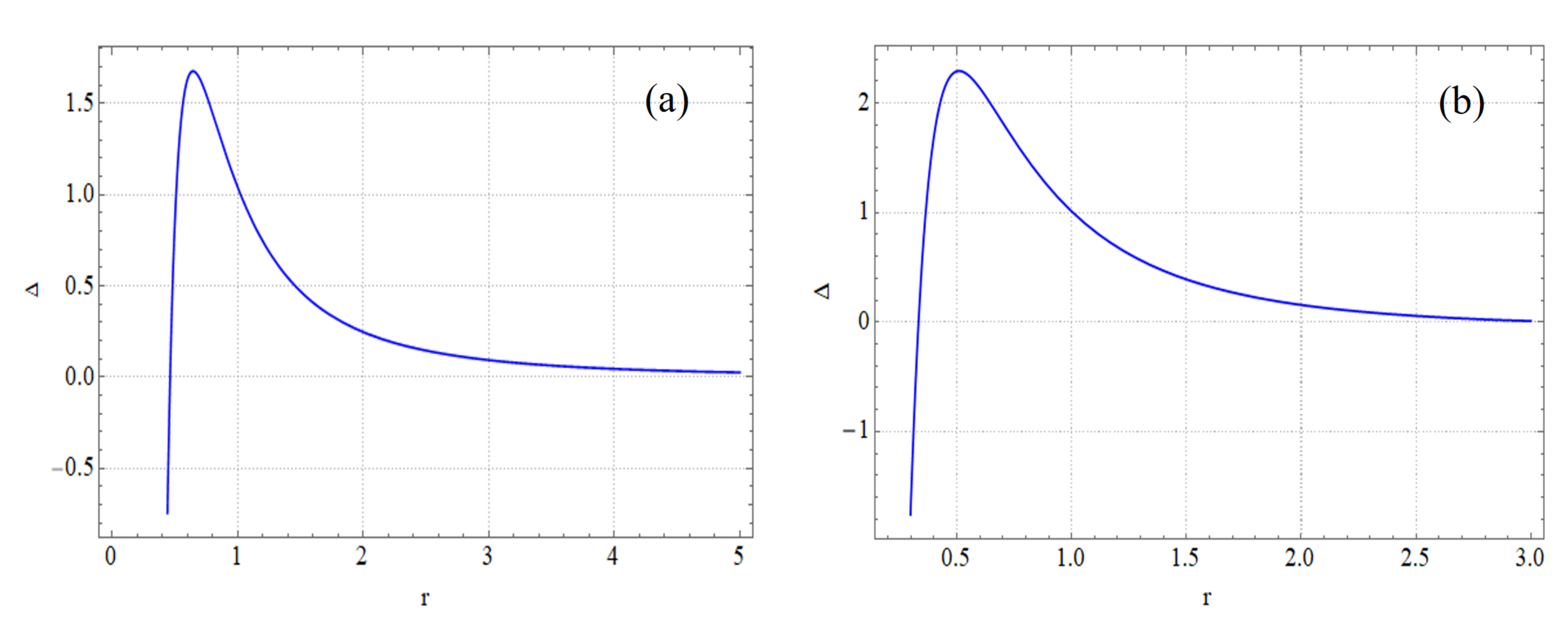}
        \caption{\centering Profile of the anisotropy parameter $\Delta$ vs. $r$ with $R_c = 10^{-29}$, \, $\mu = 0.10$, \, $p=3 \times 10^{-11}$, \, $r_{0} = 0.9$ for (a) Case I and (b) Case II.}
        \label{fig:eight}
    \end{figure}

    Figure \ref{fig:seven} shows the EoS parameters for (a) Case I and (b) Case II. It can be seen that the EoS parameter is $\omega < -1$ near the throat for both the cases, indicating a phantom like matter source threading the wormhole. Figure \ref{fig:eight} shows the anisotropy parameter $\Delta$ for (a) Case I and (b) Case II. As can be seen, the anisotropy parameter is positive near the throat, which is a characteristic feature of Morris-Thorne wormholes. Detailed analyses of these plots can be found in Table. \ref{tab:one} and Table. \ref{tab:two} respectively.

\begin{figure}[H]
        \includegraphics[width=\columnwidth]{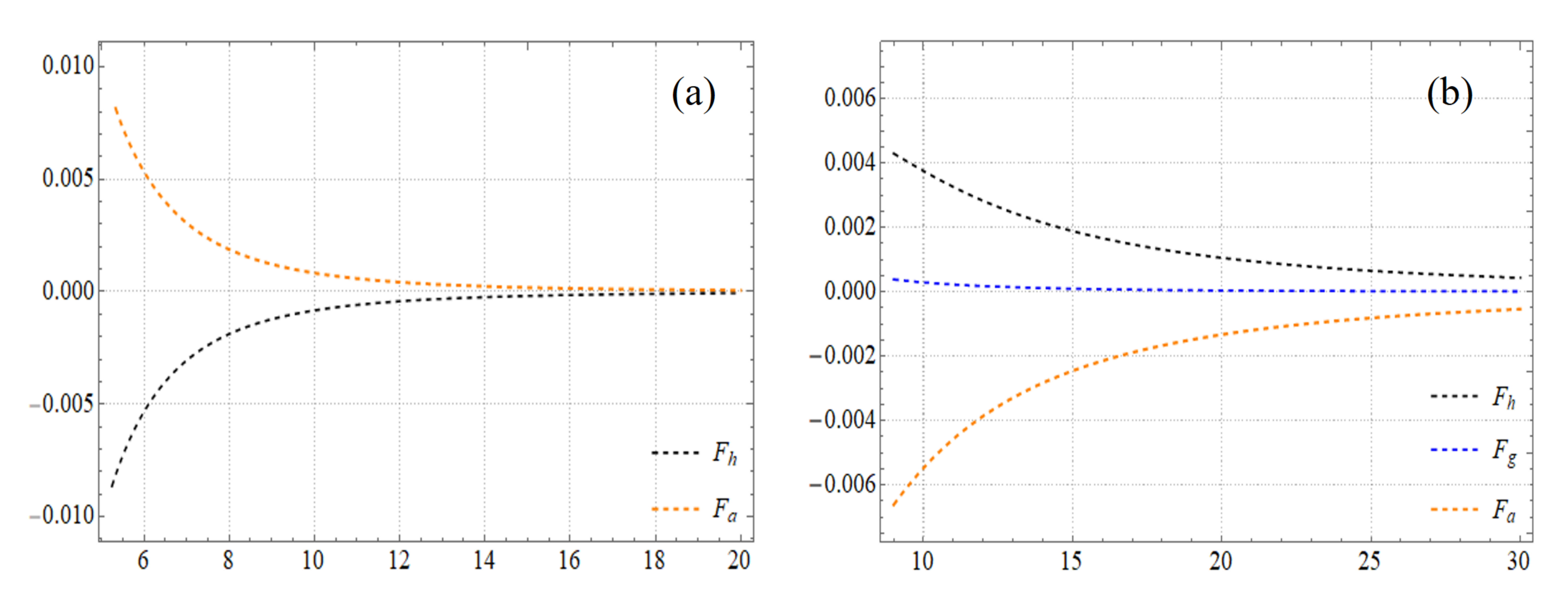}
        \caption{\centering Profile of $F_{\mathrm{h}}$ and $F_{\mathrm{a}}$ vs. $r$ for (a) Case I and $F_{\mathrm{h}}$, $F_{\mathrm{g}}$, and $F_{\mathrm{a}}$ vs. $r$ for (b) Case II with $R_c = 10^{-29}$, \, $\mu = 0.10$, \, $p=3 \times 10^{-11}$, \, $r_{0} = 0.9$.}
        \label{fig:nine}
    \end{figure}

    Next, we analyze the stability of these wormhole configurations using the TOV equation. Figure \ref{fig:nine} shows the corresponding terms of the TOV equation for (a) Case I and (b) Case II. For Case I, it is evident that the gravitational force $F_{\mathrm{g}}=0$, as $\Phi'(r)=0$. $F_{\mathrm{h}}$, the hydrostatic force and $F_{\mathrm{a}}$, the anisotropic force balance each other out, signifying a stable wormhole configuration. Similarly, Figure \ref{fig:nine}(b) shows that all the three corresponding terms of the TOV equation balance each other, rendering a stable wormhole configuration for Case II. \\ 

    \vspace{1.5cm}
    Our results in Case I (constant red-shift function) and Case II (variable red-shift function, with $\Phi(r)=\sqrt{\frac{r_0}{r}}$) demonstrate traversable wormhole solutions within the framework of a viable $f(R)$ gravity model as described in Eq. (\ref{fR}). The NEC is satisfied for both the cases. The SEC shows a oscillatory behaviour for Case I. Introducing a variable red-shift function in Case II yields a smooth plot for the SEC but it is violated at the throat. It is to be noted here that the trends of the energy condition inequalities far outside the wormhole throat have not been discussed, since the throat connecting two asymptotically flat regions is the main focus in our analyses. To this end, it is feasible to analytically cut off such solutions at some $r_c$ away from the throat, and connect it to an exterior space-time at a junction interface with a cut-off of the stress-energy at $r_c$. Considering surface stress at the junction, a thin-shell is obtained around the wormhole (see for example Ref. \cite{eiroa2016thin}), and the junction acts as a boundary surface in the absence of surface stress. Physically, this approach corresponds to wormhole space-times in which some different stress-energy distribution becomes dominant at $r>r_c$. The EoS parameter suggests a phantom like matter source threading the wormhole in both the cases. This is an interesting result in that the late-time accelerated expansion of the Universe can be ascribed to a phantom-like source of dark energy in $f(R)$ modified gravity \cite{capozziello2006cosmological}. The anisotopy parameter has a positive value near the throat signifying that the throat flares out.\\
   
\begin{figure}[H]
        \includegraphics[width=\columnwidth]{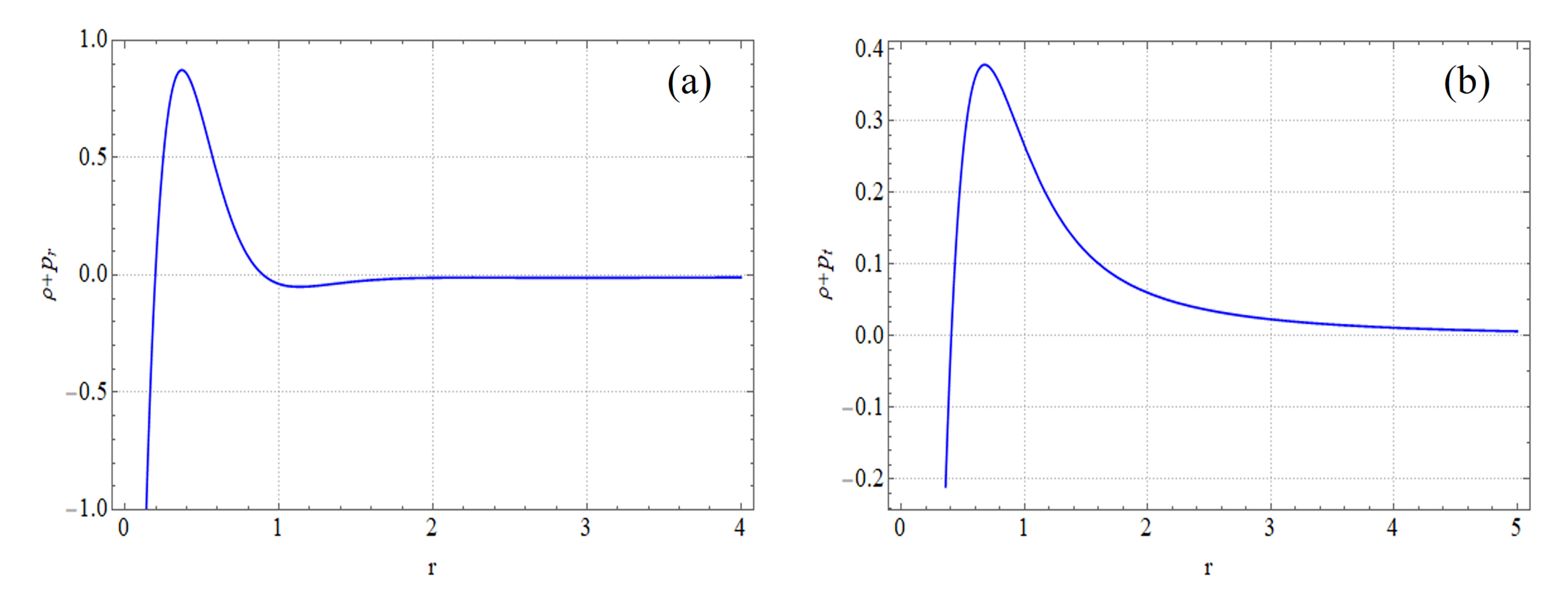}
        \caption{\centering Profile of the NEC terms (a) $\rho+p_r$ and (b) $\rho+p_t$ vs. $r$ with $M=10, \, \beta=1, \, r_{0} = 0.9$ for Case III.}
        \label{fig:ten}
    \end{figure}

    \begin{center}
    \captionof{table}{Summary of results in Case I.}
    \begin{tabular}{|c|c|c|}
        \hline
        \textbf{Term} & \textbf{Result} & \textbf{Interpretation} \\
        \hline
        $\rho$ & \makecell{$>0$ \, $\forall r$} & \makecell{WEC satisfied}  \\ 
        \hline
        $\rho+p_r$ & \makecell{$>0$ for $r \in (0,1) $,\\ $<0$ for $r \in (1,\infty)$ }  & \makecell{NEC satisfied at throat}  \\
        \hline
        $\rho+p_t$ & \makecell{$>0$ for $r \in (0.14,\infty) $,\\ $<0$ for $r \in (0,0.14)$} & \makecell{NEC satiesfied at throat} \\
        \hline
        $\rho+p_r+2p_t$ & \makecell{Oscillates} & \makecell{SEC indeterminate} \\
        \hline
        $\rho-\left|p_{r}\right|$ & \makecell{$>0$ for $r \in (0.12,0.9)$,\\ $<0$ for $r \in (0,0.12) \cup (0.9,\infty) $} & \makecell{DEC satisfied at throat} \\
        \hline
        $\rho-\left|p_{t}\right|$ & \makecell{$>0$ for $r \in (0.12,6.65)$,\\ $<0$ for $r \in (0,0.12)  \cup (6.65,\infty)$ } & \makecell{ DEC satisfied at throat} \\
        \hline
        $\omega$ & \makecell{$\omega<-1$ near throat} & \makecell{Phantom like matter source} \\
        \hline
        $\Delta$ & \makecell {$>0$ near throat} & \makecell{Repulsive geometry near throat} \\
        \hline
    \end{tabular}
    \label{tab:one}
    \end{center}
\vspace{3.5cm}
    \begin{center}
    \captionof{table}{Summary of results in Case II.}
    \begin{tabular}{|c|c|c|}
        \hline
        \textbf{Term} & \textbf{Result} & \textbf{Interpretation} \\
        \hline
        $\rho$ & \makecell{$<0$ for $r \in (0,0.2) $,\\ $>0$ for $r \in (0.2,\infty)$ } & \makecell{WEC satisfied}  \\ 
        \hline
        $\rho+p_r$ & \makecell{$>0$ for $r \in (0,0.9) $,\\ $<0$ for $r \in (0.9,\infty)$ }  & \makecell{NEC satisfied at throat}  \\
        \hline
        $\rho+p_t$ & \makecell{$>0$ \, $\forall r$} & \makecell{NEC satiesfied at throat} \\
        \hline
        $\rho+p_r+2p_t$ & \makecell{$>0$ for $r \in (0,0.45) $,\\ $<0$ for $r \in (0.45,\infty)$ } & \makecell{SEC violated} \\
        \hline
        $\rho-\left|p_{r}\right|$ & \makecell{$<0$ for $r \in (0,0.27) \cup (0.9,1.8)$,\\ $>0$ for $r \in (0.27,0.9) \cup (1.8,\infty) $} & \makecell{DEC satisfied at throat} \\
        \hline
        $\rho-\left|p_{t}\right|$ & \makecell{$<0$ for $r \in (0,0.26)$,\\ $>0$ for $r \in (0.26,\infty)$ } & \makecell{ DEC satisfied at throat} \\
        \hline
        $\omega$ & \makecell{$\omega<-1$ near throat} & \makecell{Phantom like matter source} \\
        \hline
        $\Delta$ & \makecell {$>0$ near throat} & \makecell{Repulsive geometry near throat} \\
        \hline
    \end{tabular}
    \label{tab:two}
    \end{center}
    \vspace{0.5cm}
    \noindent Next we present the results of the wormhole configuration in a non-commutative background, i.e., Case III. Figure \ref{fig:ten} shows the profile of the two NEC terms $\rho+p_r$, and $\rho+p_t$ vs. the radial distance $r$ for Case III. Figure \ref{fig:ten}(a) shows the plot of $\rho+p_r$, whereas Figure \ref{fig:ten}(b) shows the plot of $\rho+p_t$ and it can be seen that both the NEC terms are satisfied at the throat.

\begin{figure}[H]
        \includegraphics[width=\columnwidth]{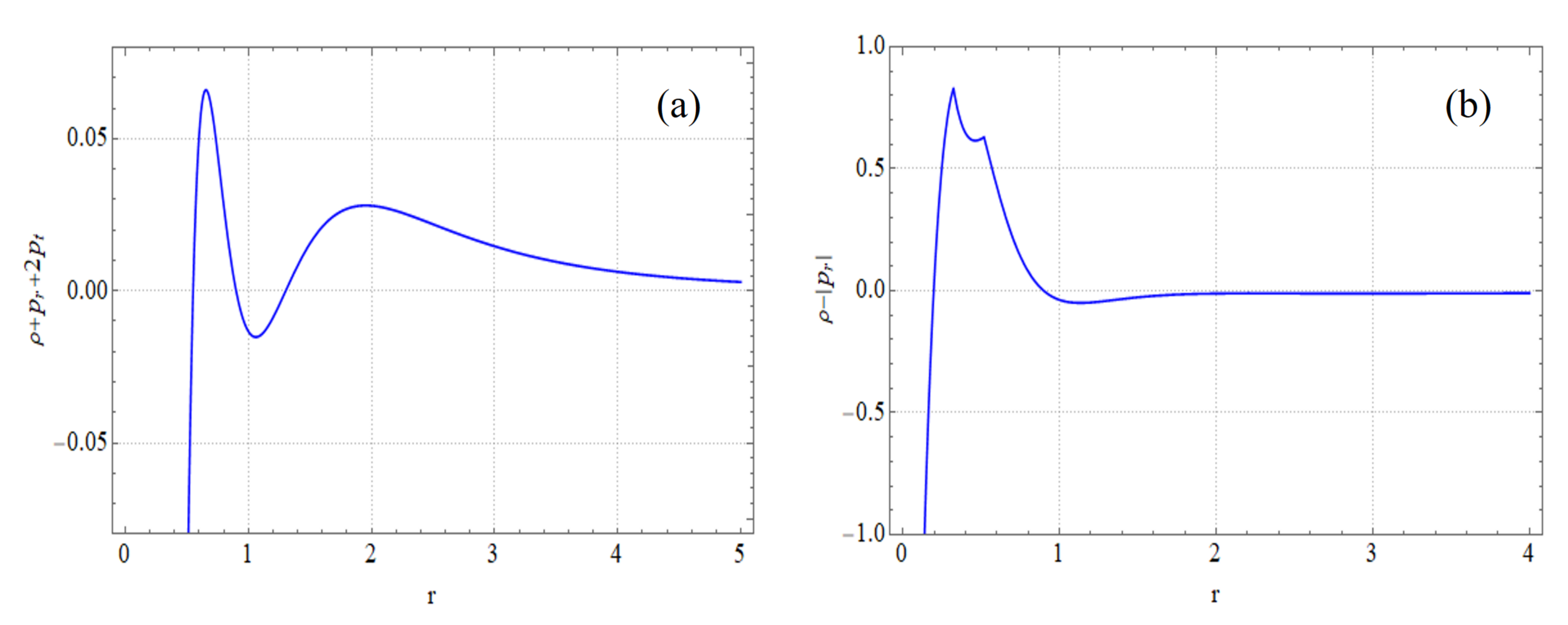}
        \caption{\centering Profile of (a) the SEC $\rho+p_r+2p_t$ and (b) DEC $\rho-|p_r|$ vs. $r$ with $M=10, \, \beta=1, \, r_{0} = 0.9$ for Case III.}
        \label{fig:eleven}
    \end{figure}

    Figure \ref{fig:eleven}(a) shows a plot of the SEC $\rho+p_r+2p_t$ vs. the radial distance $r$, and Figure \ref{fig:eleven}(b) shows the first DEC term $\rho-|p_r|$ vs. the radial distance $r$ for Case III respectively. It can be seen that both the SEC and the first DEC term are satisfied at the throat.\\

    \begin{figure}[H]
        \includegraphics[width=\columnwidth]{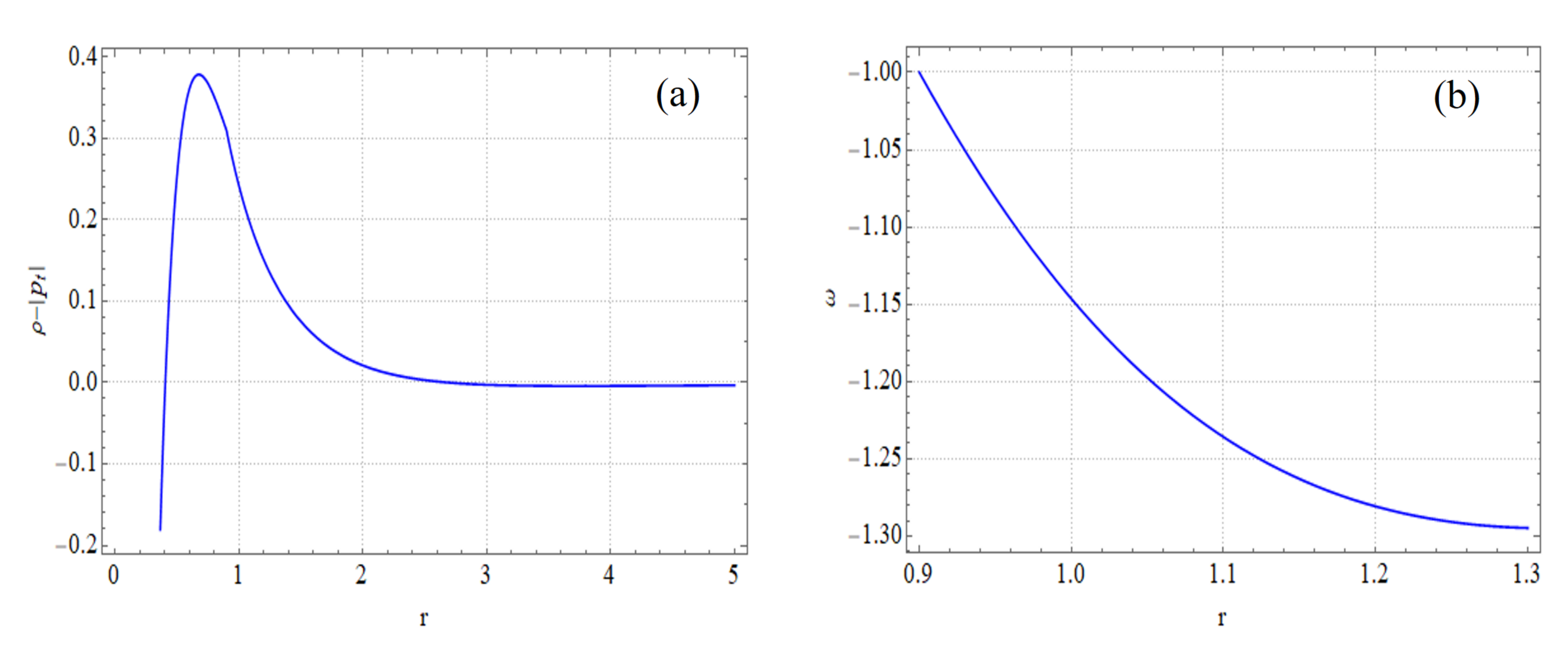}
        \caption{\centering Profile of (a) the DEC $\rho-|p_t|$ and (b) EoS parameter $\omega$ vs. $r$ with $M=10, \, \beta=1, \, r_{0} = 0.9$ for Case III.}
        \label{fig:twelve}
    \end{figure}

\begin{figure}[H]
        \includegraphics[width=\columnwidth]{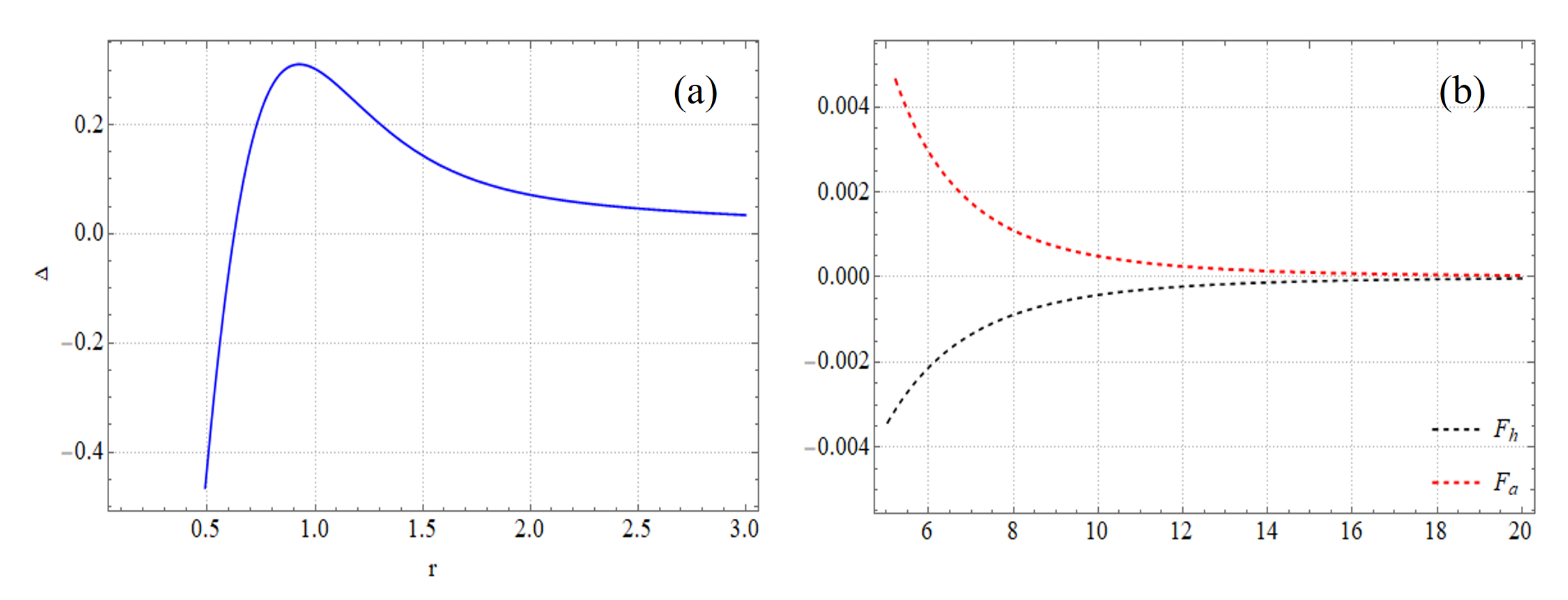}
        \caption{\centering Profile of (a) the anisotropy parameter $\Delta$ and (b) $F_{\mathrm{h}}$, and $F_{\mathrm{a}}$ vs. $r$ with $M=10, \, \beta=1, \, r_{0} = 0.9$ for Case III.}
        \label{fig:thirteen}
    \end{figure}

    Figure \ref{fig:twelve}(a) shows a plot of the second DEC term $\rho-|p_t|$ vs. the radial distance $r$ for Case III, which is satisfied. Figure \ref{fig:twelve}(b) shows the EoS parameter ($\omega$) vs. the radial distance $r$ for Case III, and $\omega < -1$ near throat, signifying a phantom like matter source. Figure \ref{fig:thirteen}(a) shows the anisotropy parameter $\Delta$ for Case III, it has a positive value near throat signifying a repulsive geometry. Figure \ref{fig:thirteen}(b) shows the corresponding terms of the TOV equation, namely, $F_{\mathrm{h}}$, and $F_{\mathrm{a}}$ balancing each other, signifying a stable wormhole configuration for Case III. Here, $F_{\mathrm{g}}=0$, as $\Phi(r)'=0$. The summary of these results can be found in Table \ref{tab:three}. \\

\begin{figure}[H]
        \includegraphics[width=\columnwidth]{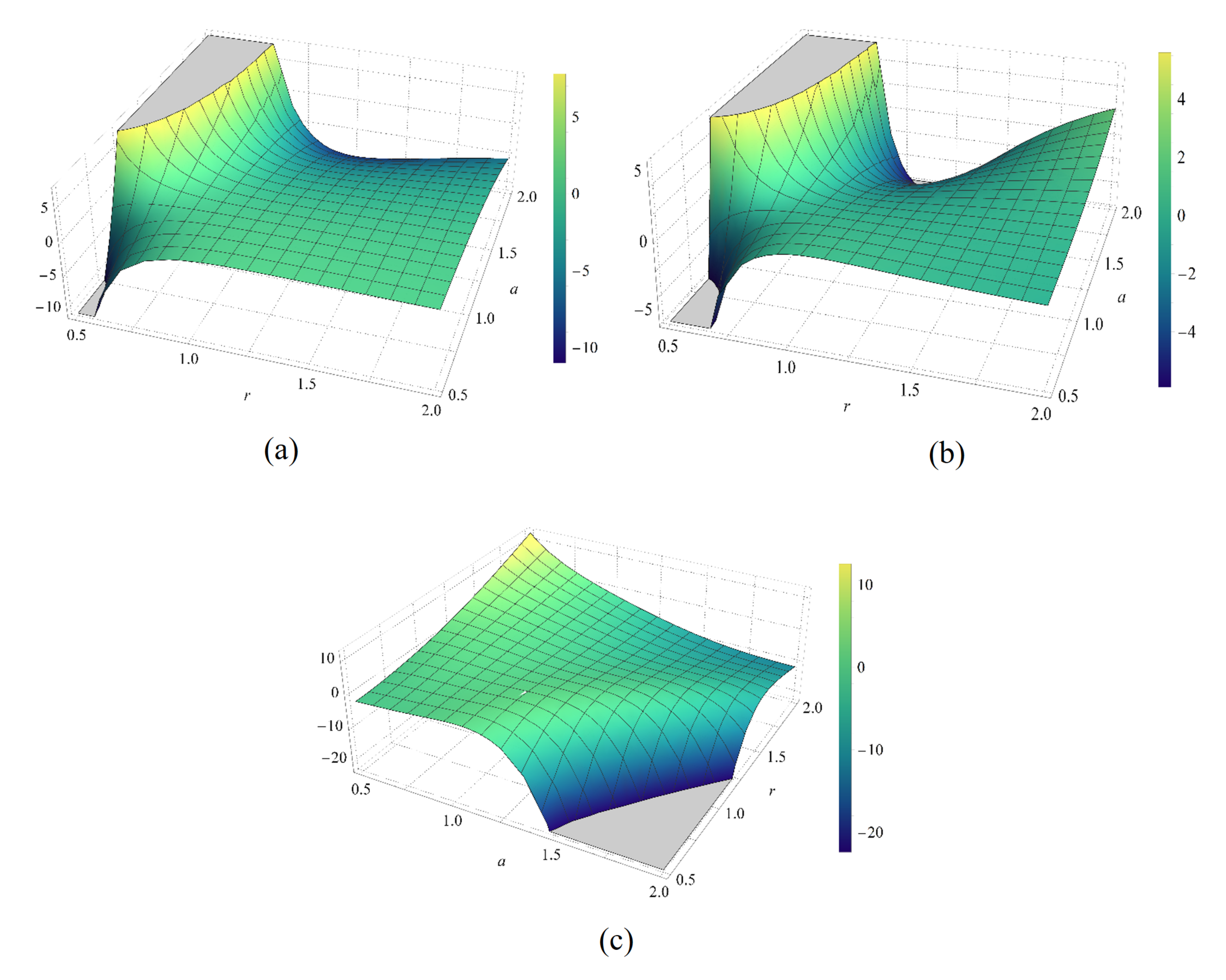}
        \caption{\centering Profile of VIQ for (a) Case I, (b) Case II, and (c) Case III}
        \label{fig:fourteen}
    \end{figure}

    \begin{center}
    \captionof{table}{Summary of results in Case III.}
    \begin{tabular}{|c|c|c|}
        \hline
        \textbf{Term} & \textbf{Result} & \textbf{Interpretation} \\
        \hline
        $\rho+p_r$ & \makecell{$>0$ for $r \in (0.2,0.9)$,\\ $<0$ for $r \in (0,0.2) \cup (0.9,\infty) $ }  & \makecell{NEC satisfied at throat}  \\
        \hline
        $\rho+p_t$ & \makecell{$>0$ for $r \in (0.41,\infty) $,\\ $<0$ for $r \in (0,0.41)$} & \makecell{NEC satisfied at throat} \\
        \hline
        $\rho+p_r+2p_t$ & \makecell{$>0$ for $r \in (0.55,0.9) \cup (1.3,\infty)$,\\ $<0$ for $r \in (0,0.55) \cup (0.9,1.3) $} & \makecell{SEC satisfied at throat} \\
        \hline
        $\rho-\left|p_{r}\right|$ & \makecell{$>0$ for $r \in (0.2,0.9)$,\\ $<0$ for $r \in (0,0.2) \cup (0.9,\infty) $} & \makecell{DEC satisfied at throat} \\
        \hline
        $\rho-\left|p_{t}\right|$ & \makecell{$>0$ for $r \in (0.41,2.7)$,\\ $<0$ for $r \in (0,0.41) \cup (2.7,\infty)$ } & \makecell{ DEC satisfied at throat} \\
        \hline
        $\omega$ & \makecell{$\omega<-1$ near throat} & \makecell{Phantom like matter source} \\
        \hline
        $\Delta$ & \makecell {$>0$ near throat} & \makecell{Repulsive geometry near throat} \\
        \hline
    \end{tabular}
    \label{tab:three}
    \end{center}
    \vspace{0.5cm}
    \noindent In Case III we invoked non-commutative geometry with the same choice of $b(r)$ as in Case I and Case II. For simplicity we assumed a constant red-shift function. Physically, this approach maybe interpreted as the brane space-time described by non-commutative operators at small length-scales, and the modified $f(R)$ theory as describing solar system and cosmological scale phenomena in four dimensions. The results in Case III demonstrate that a non-commutative background coupled with our choices of the new $f(R)$ and $b(r)$ yields a traversable wormhole satisfying the energy conditions, which is the main highlight of our results.
    \\
    Next, we analyze the volume integral quantifier to check the amount of exotic matter at the throat. As described in Section \ref{sec:whgeometry} and Eq. \ref{VIQ}, when $a \rightarrow r_0$, $I_v \rightarrow 0$  signifies that arbitrarily small quantities of NEC violating matter are required for the wormholes. We check the VIQ for all the three cases, as shown in Figure \ref{fig:fourteen}.\\

    From Figure \ref{fig:fourteen}, it can be seen that $I_v \rightarrow 0$, as $a \rightarrow r_0$ for all the three cases. It signifies that these wormhole configurations can be achieved with arbitrarily small amounts of exotic matter.
    
    \section{Conclusion}
    \label{sec:Con}
    In this study, we analyzed energy conditions for Morris-Thorne wormholes in a cosmologically viable $f(R)$ gravity model and with a non-commutative geometry background. Our results demonstrate that traversable wormholes respecting the NEC can be realized in the viable $f(R)$ model considered. Moreover, invoking a non-commutative geometry background with suitable metric parameters aids in avoiding energy condition violations. Additionally, wormholes respecting all the widely discussed energy conditions can be realized with a non-commutative geometry background.
    \\
    The study contains wormhole models with both constant and a variable redshift function and it can be seen that $\Phi(r)$ can play a crucial role in determining the energy condition inequalities. Moreover, it should be noted that the $f(R)$ model obtained via the modified EFEs in Case III may not be cosmologically viable. Eliminating the dependence of Eq. \eqref{eq:Fr} on $b(r)$ with reasonable constraints, one may test if such a scenario may yield a viable $f(R)$ model, which is another open issue. Moreover, the wormhole solutions obtained with our choices of model and metric parameters maybe investigated in terms of stability against perturbations.
    \\
    Although current observational constraints have not identified stable traversable wormholes, wormhole physics plays a crucial role in several important problems in gravity, such as the cosmic censorship conjecture \cite{penrose2002golden}, paradoxes involving closed timelike curves \cite{lobo2010closed}, the $ER=EPR$ paradigm \cite{epr}, and the nature of space-time at the smallest length scales. Studies in non-trivial space-time topologies yield wormhole solutions in most modifications of classical gravity, and also in proposed quantum approximations. Thus, the possibility of wormholes in space-time cannot be ruled out completely.

\appendix
\section{Expressions of the energy density and pressure components}\label{AppendixA}

\textbf{Case I:} The expressions for energy density $\rho$, radial pressure $p_r$, and transverse pressure $p_t$, as evaluated from Eqs. (\ref{generic1} - \ref{generic3})

\begin{align}
    \rho = \frac{r_0}{r^3} \left[1-\mu  2^{p-1} p \left(\frac{r_0}{R_c r^3}\right)^{p-1}\right]
    \end{align}
    
     \begin{align}
    p_{r}&= \frac{1}{R_c^2} \left[\mu  2^{p-3} (p-2) (p-1) p \left\{1-\frac{r_0 \log \left(\frac{r}{r_0}\right)+r_0}{r}\right\} \left(\frac{r_0}{R_c r^3}\right)^{p-3}\right] \nonumber \\
    &+ \frac{1}{r^3} \left[r_0 \left\{-\log \left(\frac{r}{r_0}\right)\right\}-r_0\right] \left[1-\mu  2^{p-1} p \left(\frac{r_0}{R_c r^3}\right)^{p-1}\right] \nonumber \\
    &+ \frac{1}{R_c r^2} \left[ \mu  2^{p-3} (p-1) p r_0 \log \left(\frac{r}{r_0}\right) \left(\frac{r_0}{R_c r^3}\right)^{p-2} \right]
    \end{align}
    
    \begin{align}
    p_{t}&= \frac{1}{R_c r} \left[ \mu  2^{p-2} (p-1) p \left\{1-\frac{r_0 \log \left(\frac{r}{r_0}\right)+r_0}{r}\right\} \left(\frac{r_0}{R_c r^3}\right)^{p-2} \right] \nonumber \\
    &+ \frac{1}{2 r^3} \left[ r_0 \log \left(\frac{r}{r_0}\right) \left\{1-\mu  2^{p-1} p \left(\frac{r_0}{R_c r^3}\right)^{p-1}\right\} \right]
    \end{align}

\textbf{Case II:} The expressions for energy density $\rho$, radial pressure $p_r$, and transverse pressure $p_t$, as evaluated from Eqs. (\ref{vrsfone} - \ref{vrsfthree})

\begingroup
\allowdisplaybreaks
\begin{align}
\rho&= \frac{(1-p) p r_0 \mu  \left(1-\frac{\log \left(\frac{r}{r_0}\right) r_0+r_0}{r}\right)}{2 R_c \sqrt{\frac{r_0}{r}} r^2} \left[\frac{1}{R_c} \left\{-\frac{\log \left(\frac{r}{r_0}\right) r_0^2}{2 \sqrt{\frac{r_0}{r}} r^4}-\frac{2 r_0}{r^3}+\left(1-\frac{\log \left(\frac{r}{r_0}\right) r_0+r_0}{r}\right) \right. \right. \nonumber \\
&\times \left. \left. \left(\frac{r_0}{r^3}-\frac{2 r_0}{\sqrt{\frac{r_0}{r}} r^3}+2 \left(\frac{r_0}{\sqrt{\frac{r_0}{r}} r^3}-\frac{r_0^2}{4 \left(\frac{r_0}{r}\right)^{3/2} r^4}\right)\right) \right\} \right]^{p-2} + \frac{1}{2} \left[1-p \mu  \left\{ \frac{1}{R_c} \left( -\frac{\log \left(\frac{r}{r_0}\right) r_0^2}{2 \sqrt{\frac{r_0}{r}} r^4} \right. \right. \right. \nonumber \\
&- \left. \left. \left. \frac{2 r_0}{r^3}+\left(1-\frac{\log \left(\frac{r}{r_0}\right) r_0+r_0}{r}\right) \left(\frac{r_0}{r^3}-\frac{2 r_0}{\sqrt{\frac{r_0}{r}} r^3}+2 \left(\frac{r_0}{\sqrt{\frac{r_0}{r}} r^3}-\frac{r_0^2}{4 \left(\frac{r_0}{r}\right)^{3/2} r^4}\right)\right) \right) \right\}^{p-1}\right] \nonumber \\
&- \frac{1}{2} \left[-\frac{\log \left(\frac{r}{r_0}\right) r_0^2}{2 \sqrt{\frac{r_0}{r}} r^4}-\frac{2 r_0}{r^3}+\left\{1-\frac{\log \left(\frac{r}{r_0}\right) r_0+r_0}{r}\right\} \left\{\frac{r_0}{r^3}-\frac{2 r_0}{\sqrt{\frac{r_0}{r}} r^3} + 2 \left(\frac{r_0}{\sqrt{\frac{r_0}{r}} r^3} \right. \right. \right. \nonumber \\
&- \left. \left. \left. \frac{r_0^2}{4 \left(\frac{r_0}{r}\right)^{3/2} r^4}\right)\right\}\right] \left[1-p \mu  \left\{ \frac{1}{R_c} \left( -\frac{\log \left(\frac{r}{r_0}\right) r_0^2}{2 \sqrt{\frac{r_0}{r}} r^4}-\frac{2 r_0}{r^3}+\left(1-\frac{\log \left(\frac{r}{r_0}\right) r_0+r_0}{r}\right) \right. \right. \right. \nonumber \\
&\times \left. \left. \left. \left(\frac{r_0}{r^3}-\frac{2 r_0}{\sqrt{\frac{r_0}{r}} r^3}+2 \left(\frac{r_0}{\sqrt{\frac{r_0}{r}} r^3}-\frac{r_0^2}{4 \left(\frac{r_0}{r}\right)^{3/2} r^4}\right)\right) \right) \right\}^{p-1}\right] - \left[1-\frac{\log \left(\frac{r}{r_0}\right) r_0+r_0}{r}\right] \nonumber \\
&\times \left[-\frac{(p-2) (p-1) p \mu}{R_c^2}  \left\{ \frac{1}{R_c} \left( -\frac{\log \left(\frac{r}{r_0}\right) r_0^2}{2 \sqrt{\frac{r_0}{r}} r^4}-\frac{2 r_0}{r^3}+\left(1-\frac{\log \left(\frac{r}{r_0}\right) r_0+r_0}{r}\right) \right. \right. \right. \nonumber \\
&\times \left. \left. \left. \left(\frac{r_0}{r^3}-\frac{2 r_0}{\sqrt{\frac{r_0}{r}} r^3}+2 \left(\frac{r_0}{\sqrt{\frac{r_0}{r}} r^3}-\frac{r_0^2}{4 \left(\frac{r_0}{r}\right)^{3/2} r^4}\right)\right) \right) \right\}^{p-3} + \frac{(p-1) p r_0 \mu}{2 R_c \sqrt{\frac{r_0}{r}} r^2} \left\{ \frac{1}{R_c} \left( -\frac{\log \left(\frac{r}{r_0}\right) r_0^2}{2 \sqrt{\frac{r_0}{r}} r^4} \right. \right. \right. \nonumber \\
&- \left. \left. \left. \frac{2 r_0}{r^3} + \left(1-\frac{\log \left(\frac{r}{r_0}\right) r_0+r_0}{r}\right) \left(\frac{r_0}{r^3}-\frac{2 r_0}{\sqrt{\frac{r_0}{r}} r^3}+2 \left(\frac{r_0}{\sqrt{\frac{r_0}{r}} r^3}-\frac{r_0^2}{4 \left(\frac{r_0}{r}\right)^{3/2} r^4}\right)\right) \right) \right\}^{p-2} \right. \nonumber \\
&+ \left. \frac{(p-1) p r_0 \mu  \log \left(\frac{r}{r_0}\right)}{2 R_c r^2 \left(1-\frac{\log \left(\frac{r}{r_0}\right) r_0+r_0}{r}\right)} \left\{ \frac{1}{R_c} \left(-\frac{\log \left(\frac{r}{r_0}\right) r_0^2}{2 \sqrt{\frac{r_0}{r}} r^4}-\frac{2 r_0}{r^3} + \left(1-\frac{\log \left(\frac{r}{r_0}\right) r_0+r_0}{r}\right) \right. \right. \right. \nonumber \\
&\times \left. \left. \left. \left(\frac{r_0}{r^3}-\frac{2 r_0}{\sqrt{\frac{r_0}{r}} r^3}+2 \left(\frac{r_0}{\sqrt{\frac{r_0}{r}} r^3}-\frac{r_0^2}{4 \left(\frac{r_0}{r}\right)^{3/2} r^4}\right)\right) \right) \right\}^{p-2} - \frac{2 (p-1) p \mu}{R_c r} \left\{ \frac{1}{R_c} \left( -\frac{\log \left(\frac{r}{r_0}\right) r_0^2}{2 \sqrt{\frac{r_0}{r}} r^4} \right. \right. \right. \nonumber \\
&- \left. \left. \left. \frac{2 r_0}{r^3}+\left(1-\frac{\log \left(\frac{r}{r_0}\right) r_0+r_0}{r}\right) \left(\frac{r_0}{r^3}-\frac{2 r_0}{\sqrt{\frac{r_0}{r}} r^3}+2 \left(\frac{r_0}{\sqrt{\frac{r_0}{r}} r^3}-\frac{r_0^2}{4 \left(\frac{r_0}{r}\right)^{3/2} r^4}\right)\right) \right) \right\}^{p-2}\right] \nonumber \\
&+\frac{r_0}{r^3}\left[1-p \mu  \left\{ \frac{1}{R_c} \left(-\frac{\log \left(\frac{r}{r_0}\right) r_0^2}{2 \sqrt{\frac{r_0}{r}} r^4}-\frac{2 r_0}{r^3}+\left(1-\frac{\log \left(\frac{r}{r_0}\right) r_0+r_0}{r}\right) \left(\frac{r_0}{r^3}-\frac{2 r_0}{\sqrt{\frac{r_0}{r}} r^3} \right. \right. \right. \right. \nonumber \\
&+ \left. \left. \left. \left. 2 \left(\frac{r_0}{\sqrt{\frac{r_0}{r}} r^3}-\frac{r_0^2}{4 \left(\frac{r_0}{r}\right)^{3/2} r^4}\right)\right)  \right) \right\}^{p-1}\right]
\end{align}
\endgroup

\begingroup
\allowdisplaybreaks
\begin{align}
p_r&=\frac{1}{2} \left[p \mu  \left\{ \frac{1}{R_c} \left( -\frac{\log \left(\frac{r}{r_0}\right) r_0^2}{2 \sqrt{\frac{r_0}{r}} r^4}-\frac{2 r_0}{r^3}+\left(1-\frac{\log \left(\frac{r}{r_0}\right) r_0+r_0}{r}\right) \left(\frac{r_0}{r^3}-\frac{2 r_0}{\sqrt{\frac{r_0}{r}} r^3} \right. \right. \right. \right. \nonumber \\
&+ \left. \left. \left. \left. 2 \left(\frac{r_0}{\sqrt{\frac{r_0}{r}} r^3}-\frac{r_0^2}{4 \left(\frac{r_0}{r}\right)^{3/2} r^4}\right)\right) \right)  \right\}^{p-1}-1\right] + \frac{1}{2} \left[1-p \mu  \left\{ \frac{1}{R_c} \left( -\frac{\log \left(\frac{r}{r_0}\right) r_0^2}{2 \sqrt{\frac{r_0}{r}} r^4}-\frac{2 r_0}{r^3} \right. \right. \right. \nonumber \\
&+ \left. \left. \left. \left(1-\frac{\log \left(\frac{r}{r_0}\right) r_0+r_0}{r}\right) \left(\frac{r_0}{r^3}-\frac{2 r_0}{\sqrt{\frac{r_0}{r}} r^3}+2 \left(\frac{r_0}{\sqrt{\frac{r_0}{r}} r^3}-\frac{r_0^2}{4 \left(\frac{r_0}{r}\right)^{3/2} r^4}\right)\right) \right) \right\}^{p-1}\right] \nonumber \\
&\times \left[-\frac{\log \left(\frac{r}{r_0}\right) r_0^2}{2 \sqrt{\frac{r_0}{r}} r^4}-\frac{2 r_0}{r^3} + \left\{1-\frac{\log \left(\frac{r}{r_0}\right) r_0+r_0}{r}\right\} \left\{\frac{r_0}{r^3}-\frac{2 r_0}{\sqrt{\frac{r_0}{r}} r^3} \right. \right. \nonumber \\
&+ \left. \left. 2 \left(\frac{r_0}{\sqrt{\frac{r_0}{r}} r^3}-\frac{r_0^2}{4 \left(\frac{r_0}{r}\right)^{3/2} r^4}\right)\right\}\right] + \left[\frac{(p-1) p r_0 \mu  \log \left(\frac{r}{r_0}\right)}{2 R_c r^2 \left(1-\frac{\log \left(\frac{r}{r_0}\right) r_0+r_0}{r}\right)} \left\{ \frac{1}{R_c} \left(-\frac{\log \left(\frac{r}{r_0}\right) r_0^2}{2 \sqrt{\frac{r_0}{r}} r^4} \right. \right. \right. \nonumber \\
&- \left. \left. \left. \frac{2 r_0}{r^3}+\left(1-\frac{\log \left(\frac{r}{r_0}\right) r_0+r_0}{r}\right) \left(\frac{r_0}{r^3}-\frac{2 r_0}{\sqrt{\frac{r_0}{r}} r^3}+2 \left(\frac{r_0}{\sqrt{\frac{r_0}{r}} r^3}-\frac{r_0^2}{4 \left(\frac{r_0}{r}\right)^{3/2} r^4}\right)\right) \right) \right\}^{p-2} \right. \nonumber \\
&- \left. \frac{(p-2) (p-1) p \mu}{R_c^2} \left\{ \frac{1}{R_c} \left(-\frac{\log \left(\frac{r}{r_0}\right) r_0^2}{2 \sqrt{\frac{r_0}{r}} r^4}-\frac{2 r_0}{r^3}+\left(1-\frac{\log \left(\frac{r}{r_0}\right) r_0+r_0}{r}\right) \right. \right. \right. \nonumber \\
&\times \left. \left. \left. \left(\frac{r_0}{r^3}-\frac{2 r_0}{\sqrt{\frac{r_0}{r}} r^3}+2 \left(\frac{r_0}{\sqrt{\frac{r_0}{r}} r^3}-\frac{r_0^2}{4 \left(\frac{r_0}{r}\right)^{3/2} r^4}\right)\right) \right) \right\}^{p-3}\right] \left[-\left(1-\frac{\log \left(\frac{r}{r_0}\right) r_0+r_0}{r}\right)\right] \nonumber \\
&+ \left[-\frac{(p-2) (p-1) p \mu}{R_c^2}  \left\{ \frac{1}{R_c} \left( -\frac{\log \left(\frac{r}{r_0}\right) r_0^2}{2 \sqrt{\frac{r_0}{r}} r^4}-\frac{2 r_0}{r^3}+\left(1-\frac{\log \left(\frac{r}{r_0}\right) r_0+r_0}{r}\right) \right. \right. \right. \nonumber \\
&\times \left. \left. \left. \left(\frac{r_0}{r^3}-\frac{2 r_0}{\sqrt{\frac{r_0}{r}} r^3}+2 \left(\frac{r_0}{\sqrt{\frac{r_0}{r}} r^3}-\frac{r_0^2}{4 \left(\frac{r_0}{r}\right)^{3/2} r^4}\right)\right) \right) \right\}^{p-3} + \frac{(p-1) p r_0 \mu}{2 R_c \sqrt{\frac{r_0}{r}} r^2} \left\{ \frac{1}{R_c} \left(-\frac{\log \left(\frac{r}{r_0}\right) r_0^2}{2 \sqrt{\frac{r_0}{r}} r^4} \right. \right. \right. \nonumber \\
&- \left. \left. \left. \frac{2 r_0}{r^3}+\left(1-\frac{\log \left(\frac{r}{r_0}\right) r_0+r_0}{r}\right) \left(\frac{r_0}{r^3}-\frac{2 r_0}{\sqrt{\frac{r_0}{r}} r^3}+2 \left(\frac{r_0}{\sqrt{\frac{r_0}{r}} r^3}-\frac{r_0^2}{4 \left(\frac{r_0}{r}\right)^{3/2} r^4}\right)\right) \right) \right\}^{p-2} \right. \nonumber \\
&+ \left. \frac{(p-1) p r_0 \mu  \log \left(\frac{r}{r_0}\right)}{2 R_c r^2 \left(1-\frac{\log \left(\frac{r}{r_0}\right) r_0+r_0}{r}\right)} \left\{ \frac{1}{R_c} \left( -\frac{\log \left(\frac{r}{r_0}\right) r_0^2}{2 \sqrt{\frac{r_0}{r}} r^4}-\frac{2 r_0}{r^3}+\left(1-\frac{\log \left(\frac{r}{r_0}\right) r_0+r_0}{r}\right) \right. \right. \right. \nonumber \\
&\times \left. \left. \left. \left(\frac{r_0}{r^3}-\frac{2 r_0}{\sqrt{\frac{r_0}{r}} r^3}+2 \left(\frac{r_0}{\sqrt{\frac{r_0}{r}} r^3}-\frac{r_0^2}{4 \left(\frac{r_0}{r}\right)^{3/2} r^4}\right)\right) \right) \right\}^{p-2} - \frac{2 (p-1) p \mu}{R_c r} \left\{ \frac{1}{R_c} \left(-\frac{\log \left(\frac{r}{r_0}\right) r_0^2}{2 \sqrt{\frac{r_0}{r}} r^4} \right. \right. \right. \nonumber \\
&- \left. \left. \left. \frac{2 r_0}{r^3}+\left(1-\frac{\log \left(\frac{r}{r_0}\right) r_0+r_0}{r}\right) \left(\frac{r_0}{r^3}-\frac{2 r_0}{\sqrt{\frac{r_0}{r}} r^3}+2 \left(\frac{r_0}{\sqrt{\frac{r_0}{r}} r^3}-\frac{r_0^2}{4 \left(\frac{r_0}{r}\right)^{3/2} r^4}\right)\right) \right) \right\}^{p-2}\right] \nonumber \\
&\times \left[1-\frac{\log \left(\frac{r}{r_0}\right) r_0+r_0}{r}\right] + \frac{\left(-\log \left(\frac{r}{r_0}\right) r_0-r_0\right)}{r^3} \left[1-p \mu  \left\{ \frac{1}{R_c} \left(-\frac{\log \left(\frac{r}{r_0}\right) r_0^2}{2 \sqrt{\frac{r_0}{r}} r^4} \right. \right. \right. \nonumber \\
&- \left. \left. \left. \frac{2 r_0}{r^3}+\left(1-\frac{\log \left(\frac{r}{r_0}\right) r_0+r_0}{r}\right) \left(\frac{r_0}{r^3}-\frac{2 r_0}{\sqrt{\frac{r_0}{r}} r^3}+2 \left(\frac{r_0}{\sqrt{\frac{r_0}{r}} r^3}-\frac{r_0^2}{4 \left(\frac{r_0}{r}\right)^{3/2} r^4}\right)\right) \right) \right\}^{p-1}\right] \nonumber \\
&+ \frac{2}{r} \left(1-\frac{\log \left(\frac{r}{r_0}\right) r_0+r_0}{r}\right) \sqrt{\frac{r_0}{r}} + \left[1-p \mu  \left\{ \frac{1}{R_c} \left(-\frac{\log \left(\frac{r}{r_0}\right) r_0^2}{2 \sqrt{\frac{r_0}{r}} r^4}-\frac{2 r_0}{r^3} \right. \right. \right. \nonumber \\
&+ \left. \left. \left. \left(1-\frac{\log \left(\frac{r}{r_0}\right) r_0+r_0}{r}\right) \left(\frac{r_0}{r^3}-\frac{2 r_0}{\sqrt{\frac{r_0}{r}} r^3} + 2 \left(\frac{r_0}{\sqrt{\frac{r_0}{r}} r^3}-\frac{r_0^2}{4 \left(\frac{r_0}{r}\right)^{3/2} r^4}\right)\right) \right) \right\}^{p-1}\right]
\end{align}
\endgroup

\begingroup
\allowdisplaybreaks
\begin{align}
p_t&=\frac{(p-1) p \left(1-\frac{\log \left(\frac{r}{r_0}\right) r_0+r_0}{r}\right) \mu}{R_c r}  \left[ \frac{1}{R_c} \left\{ -\frac{\log \left(\frac{r}{r_0}\right) r_0^2}{2 \sqrt{\frac{r_0}{r}} r^4}-\frac{2 r_0}{r^3}+\left(1-\frac{\log \left(\frac{r}{r_0}\right) r_0+r_0}{r}\right) \right. \right. \nonumber \\
&\times \left. \left. \left(\frac{r_0}{r^3}-\frac{2 r_0}{\sqrt{\frac{r_0}{r}} r^3}+2 \left(\frac{r_0}{\sqrt{\frac{r_0}{r}} r^3}-\frac{r_0^2}{4 \left(\frac{r_0}{r}\right)^{3/2} r^4}\right)\right) \right\} \right]^{p-2} + \frac{1}{2} \left[p \mu \left\{ \frac{1}{R_c} \left( -\frac{\log \left(\frac{r}{r_0}\right) r_0^2}{2 \sqrt{\frac{r_0}{r}} r^4} \right. \right. \right. \nonumber \\
&- \left. \left. \left. \frac{2 r_0}{r^3} + \left(1-\frac{\log \left(\frac{r}{r_0}\right) r_0+r_0}{r}\right) \left(\frac{r_0}{r^3}-\frac{2 r_0}{\sqrt{\frac{r_0}{r}} r^3}+2 \left(\frac{r_0}{\sqrt{\frac{r_0}{r}} r^3}-\frac{r_0^2}{4 \left(\frac{r_0}{r}\right)^{3/2} r^4}\right)\right) \right) \right\}^{p-1}-1\right] \nonumber \\
&+ \frac{1}{2} \left[1-p \mu  \left\{ \frac{1}{R_c} \left(-\frac{\log \left(\frac{r}{r_0}\right) r_0^2}{2 \sqrt{\frac{r_0}{r}} r^4}-\frac{2 r_0}{r^3}+\left(1-\frac{\log \left(\frac{r}{r_0}\right) r_0+r_0}{r}\right) \left(\frac{r_0}{r^3}-\frac{2 r_0}{\sqrt{\frac{r_0}{r}} r^3} \right. \right. \right. \right. \nonumber \\
&+ \left. \left. \left. \left. 2 \left(\frac{r_0}{\sqrt{\frac{r_0}{r}} r^3}-\frac{r_0^2}{4 \left(\frac{r_0}{r}\right)^{3/2} r^4}\right)\right) \right) \right\}^{p-1}\right] \left[-\frac{\log \left(\frac{r}{r_0}\right) r_0^2}{2 \sqrt{\frac{r_0}{r}} r^4}-\frac{2 r_0}{r^3}+\left\{1-\frac{\log \left(\frac{r}{r_0}\right) r_0+r_0}{r}\right\} \right. \nonumber \\
&\times \left. \left\{\frac{r_0}{r^3}-\frac{2 r_0}{\sqrt{\frac{r_0}{r}} r^3}+2 \left(\frac{r_0}{\sqrt{\frac{r_0}{r}} r^3}-\frac{r_0^2}{4 \left(\frac{r_0}{r}\right)^{3/2} r^4}\right)\right\}\right] + \left[-\frac{(p-2) (p-1) p \mu}{R_c^2}  \left\{ \frac{1}{R_c} \left(-\frac{\log \left(\frac{r}{r_0}\right) r_0^2}{2 \sqrt{\frac{r_0}{r}} r^4} \right. \right. \right. \nonumber \\
&- \left. \left. \left. \frac{2 r_0}{r^3} + \left(1-\frac{\log \left(\frac{r}{r_0}\right) r_0+r_0}{r}\right) \left(\frac{r_0}{r^3}-\frac{2 r_0}{\sqrt{\frac{r_0}{r}} r^3}+2 \left(\frac{r_0}{\sqrt{\frac{r_0}{r}} r^3}-\frac{r_0^2}{4 \left(\frac{r_0}{r}\right)^{3/2} r^4}\right)\right) \right) \right\}^{p-3} \right. \nonumber \\
&+ \left. \frac{(p-1) p r_0 \mu}{2 R_c \sqrt{\frac{r_0}{r}} r^2} \left\{ \frac{1}{R_c} \left(-\frac{\log \left(\frac{r}{r_0}\right) r_0^2}{2 \sqrt{\frac{r_0}{r}} r^4}-\frac{2 r_0}{r^3}+\left(1-\frac{\log \left(\frac{r}{r_0}\right) r_0+r_0}{r}\right) \left(\frac{r_0}{r^3}-\frac{2 r_0}{\sqrt{\frac{r_0}{r}} r^3} \right. \right. \right. \right. \nonumber \\
&+ \left. \left. \left. \left. 2 \left(\frac{r_0}{\sqrt{\frac{r_0}{r}} r^3}-\frac{r_0^2}{4 \left(\frac{r_0}{r}\right)^{3/2} r^4}\right)\right) \right) \right\}^{p-2} + \frac{(p-1) p r_0 \mu \log \left(\frac{r}{r_0}\right)}{2 R_c r^2 \left(1-\frac{\log \left(\frac{r}{r_0}\right) r_0+r_0}{r}\right)} \left\{ \frac{1}{R_c} \left(-\frac{\log \left(\frac{r}{r_0}\right) r_0^2}{2 \sqrt{\frac{r_0}{r}} r^4} \right. \right. \right. \nonumber \\
&- \left. \left. \left. \frac{2 r_0}{r^3} + \left(1-\frac{\log \left(\frac{r}{r_0}\right) r_0+r_0}{r}\right) \left(\frac{r_0}{r^3}-\frac{2 r_0}{\sqrt{\frac{r_0}{r}} r^3}+2 \left(\frac{r_0}{\sqrt{\frac{r_0}{r}} r^3}-\frac{r_0^2}{4 \left(\frac{r_0}{r}\right)^{3/2} r^4}\right)\right) \right) \right\}^{p-2} \right. \nonumber \\
&- \left. \frac{2 (p-1) p \mu}{R_c r} \left\{ \frac{1}{R_c} \left(-\frac{\log \left(\frac{r}{r_0}\right) r_0^2}{2 \sqrt{\frac{r_0}{r}} r^4}-\frac{2 r_0}{r^3}+\left(1-\frac{\log \left(\frac{r}{r_0}\right) r_0+r_0}{r}\right) \left(\frac{r_0}{r^3}-\frac{2 r_0}{\sqrt{\frac{r_0}{r}} r^3} \right. \right. \right. \right. \nonumber \\
&+ \left. \left. \left. \left. 2 \left(\frac{r_0}{\sqrt{\frac{r_0}{r}} r^3}-\frac{r_0^2}{4 \left(\frac{r_0}{r}\right)^{3/2} r^4}\right)\right) \right) \right\}^{p-2}\right] \left[1-\frac{\log \left(\frac{r}{r_0}\right) r_0+r_0}{r}\right] + \frac{1-p \mu r_0 \log \left(\frac{r}{r_0}\right)}{2 r^2} \nonumber \\
&\times \left[ \frac{1}{R_c} \left\{-\frac{\log \left(\frac{r}{r_0}\right) r_0^2}{2 \sqrt{\frac{r_0}{r}} r^4} - \frac{2 r_0}{r^3}+\left(1-\frac{\log \left(\frac{r}{r_0}\right) r_0+r_0}{r}\right) \left(\frac{r_0}{r^3}-\frac{2 r_0}{\sqrt{\frac{r_0}{r}} r^3} \right. \right. \right. \nonumber \\
&+ \left. \left. \left. 2 \left(\frac{r_0}{\sqrt{\frac{r_0}{r}} r^3}-\frac{r_0^2}{4 \left(\frac{r_0}{r}\right)^{3/2} r^4}\right)\right) \right\} \right]^{p-1} + \left[1-p \mu  \left\{ \frac{1}{R_c} \left(-\frac{\log \left(\frac{r}{r_0}\right) r_0^2}{2 \sqrt{\frac{r_0}{r}} r^4}-\frac{2 r_0}{r^3} \right. \right. \right. \nonumber \\
&+ \left. \left. \left. \left(1-\frac{\log \left(\frac{r}{r_0}\right) r_0+r_0}{r}\right) \left(\frac{r_0}{r^3}-\frac{2 r_0}{\sqrt{\frac{r_0}{r}} r^3} + 2 \left(\frac{r_0}{\sqrt{\frac{r_0}{r}} r^3}-\frac{r_0^2}{4 \left(\frac{r_0}{r}\right)^{3/2} r^4}\right)\right) \right) \right\}^{p-1}\right] \nonumber \\
&\times \left[1-\frac{\log \left(\frac{r}{r_0}\right) r_0+r_0}{r}\right] \left[-\frac{r_0^2}{4 \left(\frac{r_0}{r}\right)^{3/2} r^4} - \frac{\log \left(\frac{r}{r_0}\right) r_0^2}{4 \sqrt{\frac{r_0}{r}} r^3 \left(-\log \left(\frac{r}{r_0}\right) r_0-r_0+r\right)}+\frac{r_0}{2 \sqrt{\frac{r_0}{r}} r^3}\right]
\end{align}
\endgroup

\textbf{Case III:} The expressions for radial pressure $p_r$, and transverse pressure $p_t$ evaluated from Eqs. (\ref{generic2} - \ref{generic3}) are:

    \begin{align}
    p_r&=\frac{\sqrt{\beta } M \left[r_0 \left(-\log \left(\frac{r}{r_0}\right)\right)-r_0\right]}{\pi ^2 r_0 \left(\beta +r^2\right)^2}- \frac{1}{2 r^2}\left[r_0 \log \left(\frac{r}{r_0}\right) \left\{\frac{3 \sqrt{\beta } M r^2}{\pi ^2 r_0 \left(\beta +r^2\right)^2}-\frac{4 \sqrt{\beta } M r^4}{\pi ^2 r_0 \left(\beta +r^2\right)^3}\right\}\right] \nonumber \\
    &-\left[1-\frac{r_0 \log \left(\frac{r}{r_0}\right)+r_0}{r}\right] \left[\frac{6 \sqrt{\beta } M r}{\pi ^2 r_0 \left(\beta +r^2\right)^2}+\frac{24 \sqrt{\beta } M r^5}{\pi ^2 r_0 \left(\beta +r^2\right)^4} -\frac{28 \sqrt{\beta } M r^3}{\pi ^2 r_0 \left(\beta +r^2\right)^3}\right]
    \end{align}
    
    \begin{align}
    p_t&=\frac{\sqrt{\beta } M \log \left(\frac{r}{r_0}\right)}{2 \pi ^2 \left(\beta +r^2\right)^2}-\frac{1}{r} \left[1-\frac{r_0 \log \left(\frac{r}{r_0}\right)+r_0}{r}\right] \left[\frac{3 \sqrt{\beta } M r^2}{\pi ^2 r_0 \left(\beta +r^2\right)^2} - \frac{4 \sqrt{\beta } M r^4}{\pi ^2 r_0 \left(\beta +r^2\right)^3}\right]
    \end{align}
    
\section*{References}
\printbibliography[heading=none]

@book{Visser:1995cc,
      author         = "Visser, M.",
      title          = "{Lorentzian wormholes: From Einstein to Hawking}",
      publisher      = "Woodbury",
      address        = "{USA}",
      year           = "1995",
}

@article{Flamm:1916,
      author         = "Flamm, L.",
      title          = "{Republication of: Contributions to Einstein's Theory of Gravitation}",
      journal        = "Gen. Relativ. Gravit., 2015, 47, 72",
      volume         = "47",
      year           = "2015",
      pages          = "72", 
      doi            = "https://doi.org/10.1007/s10714-015-1908-2",
}

@article{Einstein:1935tc,
      author         = "Einstein, A. and Rosen, N.",
      title          = "{The Particle Problem in the General Theory of
                        Relativity}",
      journal        = "Phys. Rev.",
      volume         = "48",
      year           = "1935",
      pages          = "73-77",  
      doi            = "https://doi.org/10.1103/PhysRev.48.73"
}

@article{ellis1973ether,
  title={Ether flow through a drainhole: A particle model in general relativity},
  author={H. G. Ellis},
  journal={J. Math. Phys.},
  volume={14},
  number={1},
  pages={104--118},
  year={1973},
  publisher={American Institute of Physics},
  doi={https://doi.org/10.1063/1.1666161}
}

@article{Morris:1988cz,
      author         = "Morris, M. S. and Thorne, K. S.",
      title          = "{Wormholes in space-time and their use for interstellar
                        travel: A tool for teaching general relativity}",
      journal        = "Am. J. Phys.",
      volume         = "56",
      year           = "1988",
      pages          = "395-412",
      doi            = "https://doi.org/10.1119/1.15620"
     }

@article{PhysRevD.80.104012,
  title = {Wormhole geometries in {$f(R)$} modified theories of gravity},
  author = {F. S. N. Lobo and M. A. Oliveira},
  journal = {Phys. Rev. D},
  volume = {80},
  issue = {10},
  pages = {104012},
  numpages = {9},
  year = {2009},
  publisher = {American Physical Society},
  doi = {https://doi.org/10.1103/PhysRevD.80.104012}
}

@article{pavlovic2015wormholes,
  title={Wormholes in viable {$f(R)$} modified theories of gravity and weak energy condition},
  author={P. Pavlovic and M. Sossich},
  journal={Eur. Phys. J. C},
  volume={75},
  number={3},
  pages={1--8},
  year={2015},
  publisher={Springer},
  doi={https://doi.org/10.1140/epjc/s10052-015-3331-y}
}

@article{PhysRevD.96.044038,
  title = {Modeling wormholes in {$f(R,T)$} gravity},
  author = {P. Moraes and P. K. Sahoo},
  journal = {Phys. Rev. D},
  volume = {96},
  issue = {4},
  pages = {044038},
  numpages = {8},
  year = {2017},
  publisher = {American Physical Society},
  doi = {https://doi.org/10.1103/PhysRevD.96.044038}
}

@article{PhysRevD.97.024007,
  title = {Nonexotic matter wormholes in a trace of the energy-momentum tensor squared gravity},
  author = {P. Moraes and P. K. Sahoo},
  journal = {Phys. Rev. D},
  volume = {97},
  issue = {2},
  pages = {024007},
  numpages = {7},
  year = {2018},
  publisher = {American Physical Society},
  doi = {https://doi.org/10.1103/PhysRevD.97.024007}
}

@article{Furey_2004,
	year = {2004},
	publisher = {IOP Publishing},
	volume = {22},
	number = {2},
	pages = {313--322},
	author = {N. Furey and A. DeBenedictis},
	title = {Wormhole throats in {$R^m$} gravity},
	journal = {Class. Quantum Gravity},
	doi = {https://doi.org/10.1088/0264-9381/22/2/005}
}

@article{2019D,
   title={Non violation of energy conditions in wormholes modeling},
   volume={34},
   number={28},
   journal={Mod. Phys. Lett. A},
   publisher={World Scientific Pub Co Pte Lt},
   author={N. Godani and G. C. Samanta},
   year={2019},
   pages={1950226},
   doi={https://doi.org/10.1142/S0217732319502262}
}

@article{PhysRevD.94.044041,
  title = {Cosmological wormholes in {$f(R)$} theories of gravity},
  author = {S. Bahamonde and M. Jamil and P. Pavlovic and M. Sossich},
  journal = {Phys. Rev. D},
  volume = {94},
  issue = {4},
  pages = {044041},
  numpages = {12},
  year = {2016},
  publisher = {American Physical Society},
  doi = {https://doi.org/10.1103/PhysRevD.94.044041}
}

@article{doi:10.1142/S0218271820500686,
author = {G. C. Samanta and N. Godani and K. Bamba},
title = {Traversable wormholes with exponential shape function in modified gravity and general relativity: A comparative study},
journal = {Int. J. Mod. Phys. D},
volume = {29},
number = {09},
pages = {2050068},
year = {2020},
doi = {https://doi.org/10.1142/S0218271820500686}
}

@article{doi:10.1142/S0218271819500391,
author = {N. Godani and G. C. Samanta},
title = {Traversable wormholes and energy conditions with two different shape functions in {$f(R)$} gravity},
journal = {Int. J. Mod. Phys. D},
volume = {28},
number = {02},
pages = {1950039},
year = {2019},
doi = {https://doi.org/10.1142/S0218271819500391}
}

@article{azizi2013wormhole,
  title={Wormhole geometries in {$f(R,T)$} gravity},
  author={T. Azizi},
  journal={Int. J. Theor. Phys.},
  volume={52},
  number={10},
  pages={3486--3493},
  year={2013},
  publisher={Springer},
  doi={https://doi.org/10.1007/s10773-013-1650-z}
}

@article{shweta2020traversable,
  title={Traversable wormhole modelling with exponential and hyperbolic shape functions in  {$f(R,T)$} framework},
  author={Shweta and A. K. Mishra and U. K. Sharma},
  journal={Int. J. Mod. Phys. A},
  volume={35},
  number={25},
  pages={2050149},
  year={2020},
  publisher={World Scientific},
  doi={https://doi.org/10.1142/S0217751X20501493}
}

@article{2019A,
author = {G. C. Samanta and N. Godani},
title = {Wormhole modeling supported by non-exotic matter},
journal = {Mod. Phys. Lett. A},
volume = {34},
number = {28},
pages = {1950224},
year = {2019},
doi = {https://doi.org/10.1142/S0217732319502249}
}

@article{2020C,
   title={Wormhole modeling in {$R^2$} gravity with linear trace term},
   volume={35},
   number={08},
   journal={Int. J. Mod. Phys. A},
   publisher={World Scientific Pub Co Pte Lt},
   author={Godani, N. and Samanta, G. C.},
   year={2020},
   pages={2050045},
   doi={https://doi.org/10.1142/S0217751X20500451}
}

@article{banerjee2009topics,
  title={Topics in noncommutative geometry inspired physics},
  author={R. Banerjee and B. Chakraborty and S. Ghosh and P. Mukherjee and S. Samanta},
  journal={Found. Phys.},
  volume={39},
  number={12},
  pages={1297--1345},
  year={2009},
  publisher={Springer},
  doi={https://doi.org/10.1007/s10701-009-9349-y}
}

@article{witten1995string,
  title={String theory dynamics in various dimensions},
  author={E. Witten,},
  journal={Nucl. Phys. B.},
  volume={443},
  number={1-2},
  pages={85--126},
  year={1995},
  publisher={Elsevier}
}

@article{seiberg1999string, 
  title={String theory and noncommutative geometry},
  author={N. Seiberg and E. Witten},
  journal={J. High Energy Phys.},
  volume={1999},
  number={09},
  pages={032},
  year={1999},
  publisher={IOP Publishing},
  doi={https://doi.org/10.1088/1126-6708/1999/09/032}
}

@article{snyder1947quantized,
  title={Quantized space-time},
  author={H. S. Snyder},
  journal={Phys. Rev.},
  volume={71},
  number={1},
  pages={38},
  year={1947},
  publisher={APS},
  doi={https://doi.org/10.1103/PhysRev.71.38}
}

@article{Nicolini_2009,
	year = {2009},
	journal = {Class. Quantum Gravity},
	publisher = {{IOP} Publishing},
	volume = {27},
	number = {1},
	pages = {015010},
	author = {P. Nicolini and E. Spallucci},
	title = {Noncommutative geometry-inspired dirty black holes},
	doi = {https://doi.org/10.1088/0264-9381/27/1/015010}
}

@article{rinaldi2011new,
  title={A new approach to non-commutative inflation},
  author={M. Rinaldi},
  journal={Class. Quantum Gravity},
  volume={28},
  number={10},
  pages={105022},
  year={2011},
  publisher={IOP Publishing},
  doi={https://doi.org/10.1088/0264-9381/28/10/105022}
}

@article{NICOLINI2006547,
title = {Noncommutative geometry inspired Schwarzschild black hole},
journal = {Phys. Lett. B},
volume = {632},
number = {4},
pages = {547-551},
year = {2006},
issn = {0370-2693},
author = {P. Nicolini and A. Smailagic and E. Spallucci},
doi = {https://doi.org/10.1016/j.physletb.2005.11.004}
}

@article{rahaman2015wormhole,
  title={Wormhole inspired by non-commutative geometry},
  author={F. Rahaman and S. Karmakar and I. Karar and S. Ray},
  journal={Phys. Lett. B},
  volume={746},
  pages={73--78},
  year={2015},
  publisher={Elsevier},
  doi={https://doi.org/10.1016/j.physletb.2015.04.048}
}

@article{rahaman2012searching,
  title={Searching for higher-dimensional wormholes with noncommutative geometry},
  author={F. Rahaman and S. Islam and P. K. F Kuhfittig and S. Ray},
  journal={Phys. Rev. D},
  volume={86},
  number={10},
  pages={106010},
  year={2012},
  publisher={APS},
  doi={https://doi.org/10.1103/PhysRevD.86.106010}
}

@article{GARATTINI2009146,
title = {Self-sustained traversable wormholes in noncommutative geometry},
journal = {Phys. Lett. B},
volume = {671},
number = {1},
pages = {146-152},
year = {2009},
issn = {0370-2693},
author = {R. Garattini and F. S. N. Lobo},
doi = {https://doi.org/10.1016/j.physletb.2008.11.064}
}

@article{zubair2017existence,
  title={Existence of stable wormholes on a non-commutative-geometric background in modified gravity},
  author={M. Zubair and G. Mustafa and S. Waheed and G. Abbas},
  journal={Eur. Phys. J. C },
  volume={77},
  number={10},
  pages={1--13},
  year={2017},
  publisher={Springer},
  doi={https://doi.org/10.1140/epjc/s10052-017-5251-5}
}

@incollection{curiel2017primer,
  title={A primer on energy conditions},
  author={E. Curiel},
  booktitle={Towards a theory of spacetime theories},
  pages={43--104},
  address={New York, NY},
  year={2017},
  publisher={Springer New York},
  doi={https://doi.org/10.1007/978-1-4939-3210-8_3}
}

@article{guo2014solar,
  title={Solar system tests of {$f(R)$} gravity},
  author={J-Q. Guo},
  journal={Int. J. Mod. Phys. D},
  volume={23},
  number={04},
  pages={1450036},
  year={2014},
  publisher={World Scientific},
  doi={https://doi.org/10.1142/S0218271814500369}
}

@book{amendola2010dark,
  title={Dark energy: theory and observations},
  author={L. Amendola and S. Tsujikawa},
  year={2010},
  publisher={Cambridge University Press},
  address={Cambridge},
  doi={https://doi.org/10.1017/CBO9780511750823}
}

@article{PhysRevD.75.083504,
  title = {Conditions for the cosmological viability of {$f(R)$} dark energy models},
  author = {L. Amendola and R. Gannouji and D. Polarski and S. Tsujikawa},
  journal = {Phys. Rev. D},
  volume = {75},
  issue = {8},
  pages = {083504},
  numpages = {22},
  year = {2007},
  publisher = {American Physical Society},
  doi = {https://doi.org/10.1103/PhysRevD.75.083504}
}

@article{hamid2012entropic, 
  title={Entropic force approach to noncommutative Schwarzschild black holes signals a failure of current physical ideas},
  author={S. H. Mehdipour},
  journal={Eur. Phys. J. Plus},
  volume={127},
  number={7},
  pages={1--8},
  year={2012},
  publisher={Springer},
  doi={https://doi.org/10.1140/epjp/i2012-12080-4}
}

@article{Rahaman:2013qza,
    author = "F. Rahaman and A. Banerjee and M. Jamil and A. K. Yadav and H. Idris",
    title = "{Noncommutative Wormholes in f(R) Gravity with Lorentzian Distribution}",
    journal = "Int. J. Theor. Phys.",
    volume = "53",
    pages = "1910--1919",
    year = "2014",
    doi = "https://doi.org/10.1007/s10773-013-1993-5"
}

@article{capozziello2006cosmological,
  title={Cosmological viability of {$f(R)$} gravity as an ideal fluid and its compatibility with a matter dominated phase},
  author={S. Capozziello and S-I. Nojiri and S. D. Odintsov and A. Troisi},
  journal={Phys. Lett. B},
  volume={639},
  number={3-4},
  pages={135--143},
  year={2006},
  publisher={Elsevier},
  doi={https://doi.org/10.1016/j.physletb.2006.06.034}
}

@article{eiroa2016thin,
  title={Thin-shell wormholes with charge in {$F(R)$} gravity},
  author={E. F. Eiroa and G. F. Aguirre},
  journal={Eur. Phys. J. C},
  volume={76},
  number={3},
  pages={1--6},
  year={2016},
  publisher={Springer},
  doi={https://doi.org/10.1140/epjc/s10052-016-3984-1}
}

@article{penrose2002golden,
  title={“{Golden Oldie}”: Gravitational collapse: the role of general relativity},
  author={R. Penrose},
  journal={Gen. Relativ. Gravit.},
  volume={34},
  number={7},
  pages={1141--1165},
  year={2002},
  publisher={Citeseer},
  doi={https://doi.org/10.1023/A:1016578408204}
}

@article{lobo2010closed,
  title={Closed timelike curves and causality violation},
  author={F. S. N. Lobo},
  journal={arXiv preprint arXiv:1008.1127},
  year={2010},
  doi={https://doi.org/10.48550/arXiv.1008.1127}
}

@article{epr,
author = {L. Susskind},
title = {{ER=EPR}, {GHZ}, and the consistency of quantum measurements},
journal = {Fortschritte der Phys.},
volume = {64},
number = {1},
pages = {72-83},
year = {2016},
doi = {https://doi.org/10.1002/prop.201500094}
}

@article{oppenheimer1939massive,
  title={On massive neutron cores},
  author={R. J. Oppenheimer and G. M. Volkoff},
  journal={Physical Review},
  volume={55},
  number={4},
  pages={374},
  year={1939},
  doi={https://doi.org/10.1103/PhysRev.55.374}
}

@article{gorini2008tolman,
  title={Tolman-Oppenheimer-Volkoff equations in the presence of the Chaplygin gas: Stars and wormholelike solutions},
  author={Gorini, V and Moschella, U and Kamenshchik, A Yu and Pasquier, V and Starobinsky, AA},
  journal={Physical Review D},
  volume={78},
  number={6},
  pages={064064},
  year={2008},
  doi={https://doi.org/10.1103/PhysRevD.78.064064}
}

@article{Kuhfittig:2020fue,
    author = {Kuhfittig, P. K. F.},
    title = {A note on the stability of Morris-Thorne wormholes},
    journal = {Fund. J. Mod. Phys.},
    volume = {14},
    pages = {23--31},
    year = {2020},
    doi={https://doi.org/10.48550/arXiv.2009.11179}
}

@article{ponce1993limiting,
  title={Limiting configurations allowed by the energy conditions},
  author={J. Ponce de Leon},
  journal={General relativity and gravitation},
  volume={25},
  number={11},
  pages={1123--1137},
  year={1993},
  doi={https://doi.org/10.1007/BF00763756}
}

@article{visser2003traversable,
  title={Traversable wormholes with arbitrarily small energy condition violations},
  author={M. Visser and S. Kar and N. Dadhich},
  journal={Physical Review Letters},
  volume={90},
  number={20},
  pages={201102},
  year={2003},
  doi={https://doi.org/10.1103/PhysRevLett.90.201102}
}

@article{kar2004quantifying,
  title={Quantifying energy condition violations in traversable wormholes},
  author={S. Kar, and N. Dadhich and M. Visser},
  journal={Pramana},
  volume={63},
  number={4},
  pages={859--864},
  year={2004},
  doi={https://doi.org/10.1007/BF02705207}
}

@article{lobo2013new,
  title={New asymptotically flat phantom wormhole solutions},
  author={F. S. N. Lobo and F. Parsaei and N. Riazi},
  journal={Physical Review D},
  volume={87},
  number={8},
  pages={084030},
  year={2013},
  doi={https://doi.org/10.1103/PhysRevD.87.084030}
}

@article{capo1,
  title = {Generalized energy conditions in extended theories of gravity},
  author = {S. Capozziello, F.S.N. Lobo, and J.P. Mimoso},
  journal = {Physical Review D},
  volume = {91},
  issue = {12},
  pages = {124019},
  year = {2015},
  doi = {https://doi.org/10.1103/PhysRevD.91.124019}
}

@article{capo2,
title = {Energy conditions in modified gravity},
author = {S. Capozziello, F.S.N. Lobo, and J.P. Mimoso},
journal = {Physics Letters B},
volume = {730},
pages = {280-283},
year = {2014},
doi = {https://doi.org/10.1016/j.physletb.2014.01.066}
}

@article{moraes2017simplest,
  title={The simplest non-minimal matter--geometry coupling in the f (R, T) cosmology},
  author={P.H.R.S Moraes, and P.K. Sahoo},
  journal={European Physical Journal C},
  volume={77},
  number={7},
  pages={1--8},
  year={2017},
  doi={https://doi.org/10.1140/epjc/s10052-017-5062-8}
}

@article{albareti2013non,
  title={On the non-attractive character of gravity in f (R) theories},
  author={F.D. Albareti and J.A.R. Cembranos and A. de la Cruz-Dombriz and A. Dobado},
  journal={Journal of Cosmology and Astroparticle Physics},
  volume={2013},
  number={07},
  pages={009},
  year={2013},
  doi={https://doi.org/10.1088/1475-7516/2013/07/009}
}

@article{fq,
  title={Wormhole solutions in symmetric teleparallel gravity with noncommutative geometry},
  author={Z. Hassan and G. Mustafa and P.K. Sahoo},
  journal={Symmetry},
  volume={13},
  number={7},
  pages={1260},
  year={2021},
  doi={https://doi.org/10.3390/sym13071260}
}

@article{fq2,
title = {Traversable wormholes with charge and non-commutative geometry in the f(Q) gravity},
journal = {Annals of Physics},
volume = {443},
pages = {168968},
year = {2022},
doi = {https://doi.org/10.1016/j.aop.2022.168968},
author = {O. Sokoliuk and Z. Hassan and P.K. Sahoo and A. Baransky}
}

@article{fq3,
title = {Wormhole solutions in symmetric teleparallel gravity},
journal = {Physics Letters B},
volume = {821},
pages = {136612},
year = {2021},
doi = {https://doi.org/10.1016/j.physletb.2021.136612},
author = {G. Mustafa and Zinnat Hassan and P.H.R.S. Moraes and P.K. Sahoo}
}

@article{fq4,
title = {Traversable wormhole inspired by non-commutative geometries in f(Q) gravity with conformal symmetry},
journal = {Annals of Physics},
volume = {437},
pages = {168751},
year = {2022},
doi = {https://doi.org/10.1016/j.aop.2021.168751},
author = {G. Mustafa and Z. Hassan and P.K. Sahoo}
}

@article{ft1,
  title={Wormholes in a viable f (T) gravity},
  author={M. Jamil and D. Momeni and R. Myrzakulov},
  journal={The European Physical Journal C},
  volume={73},
  number={1},
  pages={1--13},
  year={2013},
  doi={https://doi.org/10.1140/epjc/s10052-012-2267-8}
}

@article{ft2,
  title={Wormhole solutions in f (T) gravity with noncommutative geometry},
  author={M. Sharif and S. Rani},
  journal={Physical Review D},
  volume={88},
  number={12},
  pages={123501},
  year={2013},
  doi={https://doi.org/10.1103/PhysRevD.88.123501}
}

@article{ft3,
  title={Wormhole geometries in modified teleparallel gravity and the energy conditions},
  author={C.G. Boehmer and T. Harko and F.S.N Lobo},
  journal={Physical Review D},
  volume={85},
  number={4},
  pages={044033},
  year={2012},
  doi={https://doi.org/10.1103/PhysRevD.85.044033}
}

\end{document}